\documentclass[%
 reprint,
superscriptaddress,
 amsmath,amssymb,
 aps,
prb,
]{revtex4-2}

\usepackage{xcolor}
\usepackage{graphicx}
\usepackage{dcolumn}
\usepackage{bm}
\usepackage{placeins}
\usepackage{appendix}
\usepackage[utf8]{inputenc}
\usepackage[english]{babel}
\usepackage{float}

\begin{document}

\preprint{APS/123-QED}

\title{Emergent magnetic behaviour in the frustrated  Yb$_3$Ga$_5$O$_{12}$ garnet.}

\author{Lise $\O$rduk Sandberg}
\affiliation{Nanoscience Center, Niels Bohr Institute, University of Copenhagen, Universitetsparken 5, DK-2100 Copenhagen \O , Denmark}

\author{Richard Edberg}
\affiliation{
Physics Department, KTH Royal Institute of Technology, Sweden
}

\author{Ingrid-Marie Berg Bakke}
\affiliation{
University of Oslo, Centre for Materials Science and Nanotechnology, NO-0315, Oslo, Norway
}

\author{Kasper S. Pedersen}
\affiliation{
Department of Chemistry, Technical University of Denmark, DK-2800 Kgs. Lyngby, Denmark
}%
\author{Monica Ciomaga Hatnean}
\author{Geetha Balakrishnan}
\affiliation{
Department of Physics, University of Warwick, Coventry CV4 7AL, UK
}%

\author{Lucile Mangin-Thro}
\author{Andrew Wildes}
\author{B. Fåk}
\affiliation{
Institut Laue-Langevin, 71 Avenue des Martyrs, CS 20156, 38042 Grenoble Cedex 9, France
}
\author{Georg Ehlers}
\author{Gabriele Sala}
\affiliation{
Neutron Technologies Division, Oak Ridge National Laboratory, Oak Ridge, TN 37831-6466, USA}%

\author{Patrik Henelius}
\affiliation{
Physics Department, KTH Royal Institute of Technology, Sweden
}
\affiliation{Faculty of Science and Engineering,  \r{A}bo Akademi University, \r{A}bo, Finland}

\author{Kim Lefmann}
\affiliation{Nanoscience Center, Niels Bohr Institute, University of Copenhagen, Universitetsparken 5, DK-2100 Copenhagen \O , Denmark}

\author{Pascale P. Deen}

\affiliation{Nanoscience Center, Niels Bohr Institute, University of Copenhagen, Universitetsparken 5, DK-2100 Copenhagen \O , Denmark}
\affiliation{European Spallation Source ERIC, 22363 Lund, Sweden}
\affiliation{Corresponding author: pascale.deen@ess.eu}
\date{\today}

\begin{abstract}
We report neutron scattering, magnetic susceptibility and Monte Carlo theoretical analysis to verify the short range nature of the magnetic structure and spin-spin correlations in a  Yb$_3$Ga$_5$O$_{12}$ single crystal. The quantum spin state of Yb$^{3+}$ in Yb$_3$Ga$_5$O$_{12}$ is verified.
The quantum spins organise into a short ranged emergent director state for T $<$ 0.6 K derived from anisotropy and near neighbour exchange. We derive the magnitude of the near neighbour exchange interactions $0.6\; {\rm K} < J_1 < 0.7\; {\rm K}, J_2 = 0.12$~K and the magnitude of the dipolar exchange interaction, $D$, in the range $0.18 < D < 0.21$ K. Certain aspects of the broad experimental dataset can be modelled using a $J_1D$ model with ferromagnetic near neighbour spin-spin correlations while other aspects of the data can be accurately reproduced using a $J_1J_2D$ model with antiferromagnetic near neighbour spin-spin correlation. As such, although we do not quantify all the relevant exchange interactions we nevertheless provide a strong basis for the understanding of the complex Hamiltonian required to fully describe the magnetic state of Yb$_3$Ga$_5$O$_{12}$.  
\end{abstract}

\maketitle


\section{Introduction} \label{sec:intro}


In recent years, emergent behaviour has been observed in 3 dimensional (3D) geometrically frustrated compounds, due to the interplay between spin-spin interactions and anisotropy. In spin-ice compounds Ho$_2$Ti$_2$O$_7$ (HTO) and Dy$_2$Ti$_2$O$_7$ (DTO), with magnetic rare earth ions placed on the 3D pyrochlore lattice, a strongly correlated ground state is observed with remarkable excitations that can be modelled as magnetic monopoles. This new physics is derived from a combination of ferromagnetic (FM) nearest neighbour (NN) spin-spin interactions and a strong local Ising anisotropy along the central axes of each tetrahedron \cite{Gardner2010,Fennell2009,Morris_2017,Castelnovo2012}. 

\begin{figure}[!h]
    	\centering
    	\includegraphics[width = 0.4\linewidth]{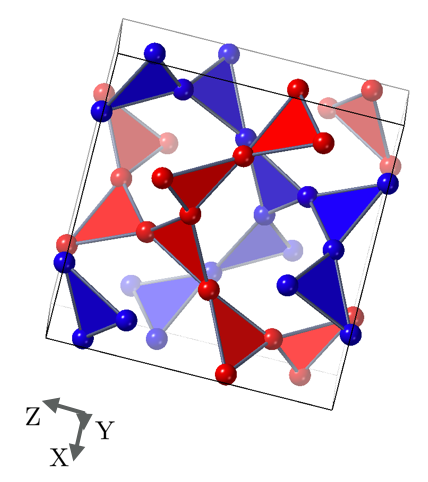}
    	\includegraphics[width = 0.5\linewidth]{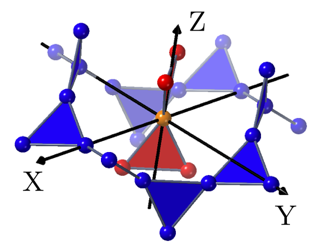}    			
		\caption{Left: 24 Yb$^{3+}$ ions in a unit cell of YbGG. Blue and red atoms are Yb ions of the two interpenetrating hyperkagome lattices respectively. Triangle surfaces between neighbouring Yb ions are coloured. Right: Local coordinate system of the central orange ion, which is located in the center of the blue 10-ion loop.  }
        \label{fig:YbGG}
\end{figure}

A second emergent state, that has recently come to light, is the long range multipolar director state found in the 3D hyperkagome structure 
Gd$_3$Ga$_5$O$_{12}$ (GGG) \cite{Paddison_2015_GGG}. In GGG the Gd$^{3+}$ ions are positioned on two interpenetrating hyperkagome lattices, shown in Fig.~\ref{fig:YbGG}. Despite the absence of long range correlations of the individual spins, an emergent long range hidden order known as a director state has been determined. The director state is derived from the collective spins on a 10-ion loop and is defined as 
\begin{equation}
    \textbf{L}(\textbf{rc})=\frac{1}{10}\sum_n \cos(n\pi) \textbf{S}_n(\textbf{r}),  \label{eq:director}
\end{equation}
where $\textbf{S}_n(\textbf{r})$ are unit-length spins on the the ten-ion loop with center in $\textbf{rc}$. The director state was found to display long range correlations in GGG and governs both the magnetic structure \cite{Paddison_2015_GGG} and magnetic dynamics \cite{Ambrumenil2015} into the high field regime. The director state is derived from anisotropy and near neighbour exchange. Gd$^{3+}$ ions display a nominal zero orbital angular momentum $L=0$ and thus no strong anisotropy due to spin-orbit coupling. However, the spins in GGG are highly anisotropic in the local XY-plane, defined in Fig.~\ref{fig:YbGG}. This anisotropy could be derived from the dipole exchange interaction and, along with antiferromagnetic (AFM) near neighbour (NN) interactions, is essential for the formation of the director state. Furthermore, as the temperature is reduced below $T < 0.175$~K, GGG enters a spin slush state, a coexistence of longer-range, solid like, and shorter range, liquid-like, correlations \cite{Petrenko1998}, that has theoretically been shown to require the inclusion of the very long range nature of the dipolar interactions  \cite{Rau_2016} and inter-hyperkagome exchange. 

The director and spin slush states in GGG can be contrasted with the unusual long range magnetic structures observed in the isostructural compounds Tb$_3$Gd$_5$O$_{12}$ (TGG) and Er$_3$Al$_5$O$_{12}$ (ErAG), \cite{TbGG_2019},\cite{ErGG_2019} for $T \leq T_{\rm N} = 0.25$~K and 0.8~K, respectively. Both compounds reveal strong local anisotropy resulting in an ordered multi-axis AFM ground state. The ground state in both compounds has been ascribed to the interaction between local anisotropy and long range dipolar interactions. The effect of dipolar interactions on Ising spins on the garnet lattice has been investigated by Monte Carlo simulations revealing a variety of distinct phases with the phase diagram strongly affected by the cut-off length of the long range interactions \cite{Yoshioka2004}.

The diverse states of matter observed in these 3D compounds depend on the perturbative effect of the anisotropy on the exchange interactions as the rare earth ions are exchanged in the hyperkagome structure. As such, we now study Yb$_3$Ga$_5$O$_{12}$ (YbGG). Significant spin–orbit interaction from the ground level $^2$F$_{7/2}$ of the Yb$^{3+}$ ions provides strong anisotropy. The YbGG room temperature unit cell lattice parameter, $a= 12.204(4)$~\AA, smaller than GGG ($a = 12.385$~\AA), ErGG ($a = 12.265$~\AA ) \cite{ErGG_2019} and TbGG ($a = 12.352$~\AA )\cite{TbGG_2019} , will affect the dipole exchange interaction. YbGG also presents the possibility to study quantum effects via the effective $S=1/2$ state due to the effect of the crystal field that acts on the Yb$^{3+}$ $^2$F$_{7/2}$  state to leave a ground state Kramers doublet, well isolated from a series of excited Kramers doublets \cite{Pearson_1967}. It is widely expected that quantum effects on a 3D frustrated lattice could lead to novel states of matter including a quantum spin liquid state, topological order and quantum entanglement \cite{Broholm2020}. 

Previously, heat capacity and magnetic susceptibility measurements on YbGG revealed a lambda transition at 0.054 K in addition to a broad peak centered at 0.18 K that extends to 0.6 K, \cite{Filippi_1980}. The energy scale of the interactions, extracted by a Curie-Weiss fit, yields $\theta_{\text{CW}}$ = 0.045(5) K, showing dominant FM interactions \cite{Filippi_1980}. The lambda transition was assigned to an ordered magnetic state, however this is not confirmed by muon spin resonance and M\"ossbauer spectroscopy from which a disordered moment has been determined down to 0.036 K, \cite{Dalmas_2003,Hodges_2003}. The broad peak centered at 0.18 K resembles the specific heat anomaly in GGG indicative of the correlated director state \cite{Paddison_2015_GGG}.  

Here, single crystal studies on YbGG are presented. We have employed neutron scattering techniques, magnetic susceptibility and Monte Carlo theoretical analysis to verify the short range nature of the magnetic structure and spin-spin correlations in YbGG.

\section{Method} \label{sec:methods}
\subsection{Experimental Method}
A single crystal of YbGG was grown using the floating zone method in Ar + O$_2$ gas atmosphere at a growth rate of 10 mm/h\; \cite{Kimura_2009}. X-ray Laue diffraction was used to determine the quality of the crystal and to align the samples used for the magnetic properties measurements.

Susceptibility measurements were performed for 1.8~K $< T <$  300~K at the Technical University of Denmark on a 0.29~g YbGG single crystal using the VSM and AC-MSII options on a Quantum Design Dynacool PPMS. Cold and thermal inelastic neutron spectroscopy and polarised neutron diffraction have been performed on a 1.9~g YbGG single crystal to access the spin-spin correlations and crystal field levels \cite{Squires}. 

Cold neutron spectroscopy was performed at the time-of-flight cold neutron chopper spectrometer (CNCS) at the Spallation Neutron Source, Oak Ridge National laboratory \cite{CNCS_Overview}. Measurements were performed at 0.05~K with incident neutron energies $E_{\rm i} = 1.55$~meV and 3.32~meV. The energy resolutions, obtained via the incoherent scattering of a Vanadium sample, are $\Delta E_i = 0.0371(5)$ and 0.109(2) meV, respectively, while the Q-resolutions were significantly narrower than the observed features\cite{CNCS_Overview}. The scattering plane comprises (-H,~H,~0) and (L,~L,~2L) with the sample rotated through 180$^\circ$ using 2$^\circ$ steps in order to access a complete rotational plane. 

Polarised neutron diffraction was performed using the diffuse scattering spectrometer D7, at the Institut Laue-Langevin (ILL), Grenoble, \cite{D7_Overview}, with E$_{\rm i}$ = 8.11 meV and a sample temperature of 0.05~K, \cite{D7_Experiment}. D7 provides an energy integrated measurement. The scattering plane again comprises of (-H,~H,~0) and (L,~L,~2L) with the sample rotating through 180$^{\circ}$ using 1$^{\circ}$ steps. D7 also provides a Q-resolution that is significantly narrower than the observed features \cite{Fennell_2017}. Calibration for detector and polarisation efficiency have been performed using Vanadium and Quartz, respectively. An empty can measurement at 50~K provides a background subtraction for non-sample dependent scattering.  

The experimental temperature determined on CNCS and D7 were stable and experimentally determined to be 0.05~K, yet the long range order expected below the lambda transition of 0.054~K was not observed. Rare earth garnets compounds display very low thermal conductivity, in particular at low temperatures. In addition, it is possible that a poor thermal contact between the sample and the thermal bath leads to higher temperatures than provided by thermometry. The specific heat measurements indicate a short ranged broad feature for T $<$ 0.6 K preceding the lambda transition. We believe both the D7 and CNCS datasets to probe the short ranged ordered regime, $0.07 $~K~$< T < 0.6$~K since the correlation lengths of the magnetic scattering is short ranged, see \ref{sec:results:scattering}.

Thermal inelastic neutron scattering measurements have been performed at the ILL to access the crystal field levels. We employed the thermal time-of-flight spectrometer, IN4, with an incident energy $E_{\rm i}$~=~113~meV at a temperature of 1.5~K. Measurements were performed for three different sample orientations with no observed angular dependence, \cite{IN4_Experiment}. YbGG crystal field parameters were extracted using the combined data. 

\subsection{Analysis Method}\label{sec:methods:analysis}
We have modelled the elastic neutron scattering profiles using the Reverse Monte Carlo (RMC) \textit{Spinvert} refinement program \cite{Paddison_2013_RMC}. The algorithm employs simulated annealing to determine real space correlations from the neutron scattering data. We simulate cubic supercells with side $L\in[1,8]$ unit cells, corresponding to a maximum number of $24 \cdot 8^3 = 12288$ spins. To obtain good statistical accuracy, we performed up to $400$ refinements and employed an average of these to derive the final correlations. In order to aid visualization we employed an interpolation technique frequently used in the Spinvert program package, \textit{windowed-sinc filtering}\cite{Paddison_2013_RMC}. The interpolation allows us to calculate $S(\textbf{Q})$ at a wavevector transfer that is not periodic in the supercell \cite{Paddison_2013_RMC}.

The RMC simulations yield information on the spin correlations, but not on the magnitude of the interactions. In order to obtain information on the interaction strengths, we have performed Monte Carlo (MC) simulations of an Ising system with nearest, next to nearest and long range dipolar interactions. The crude Ising approximation is motivated on two fronts: 
\newline 1. The heat capacity measured by Filippi et al.\cite{Filippi_1980} shows qualitative resemblance with that of a long range dipolar Ising model\cite{Yoshioka2004}. 
\newline 2. The resultant correlations from the RMC (Spinvert) algorithm, suggest that there is an easy axis along the local $z$-direction.
We have optimised the interaction parameters in the MC simulation to match the experimentally observed heat capacity. From the interactions we have computed $S(\textbf{Q})$ scattering profiles to see how they compare with the experimentally observed scattering profile, $S(\textbf{Q})$. We employed Ewald summation to handle the conditionally convergent dipolar sum.  

\section{Experimental Results} \label{sec:results}
\subsection{Susceptibility}\label{sec:results:chi}
\begin{figure}[!h]
    	\centering
        \includegraphics[width =  \linewidth]{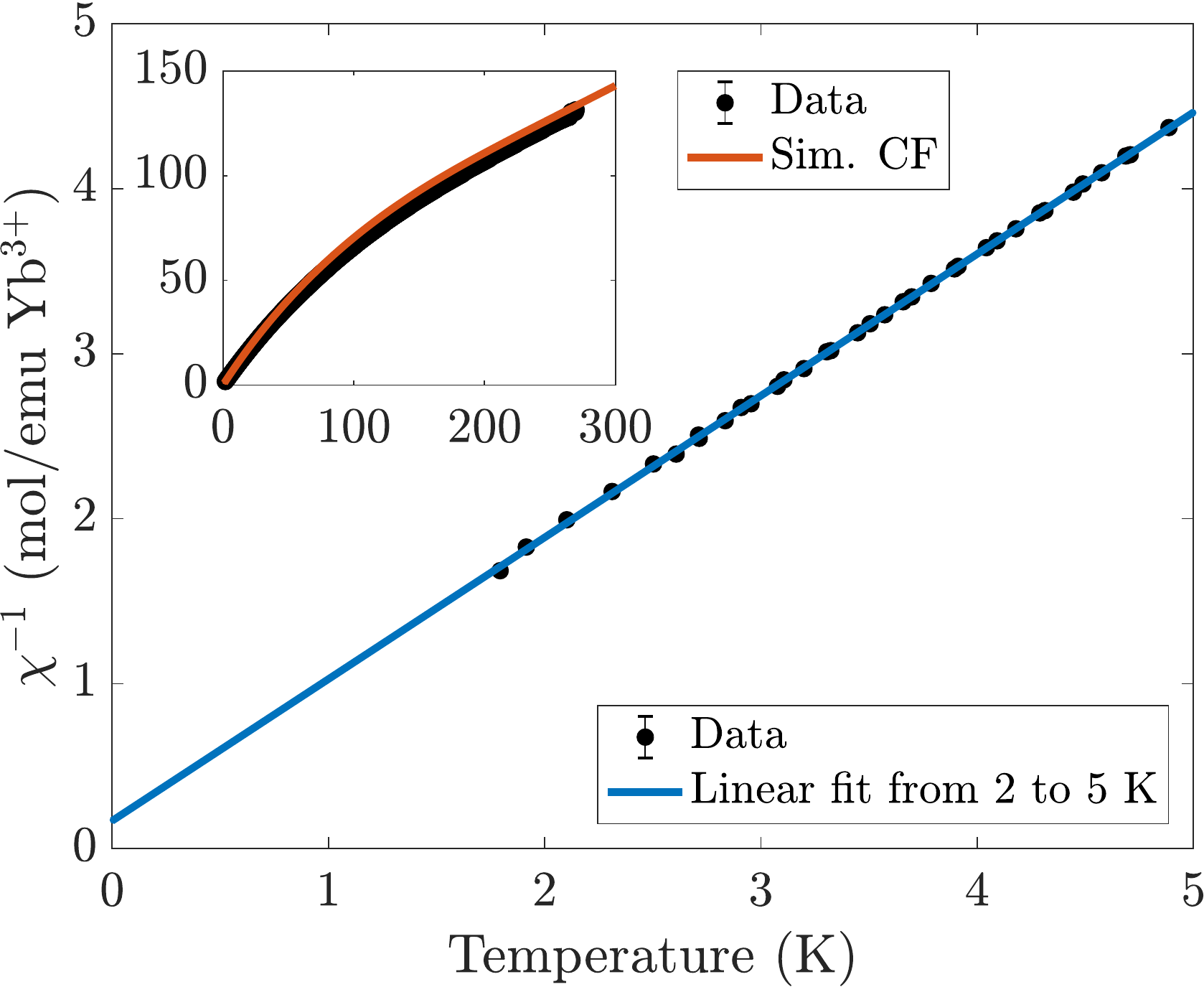}
		\caption{Inverse susceptibility from PPMS measurements of single crystal YbGG and a linear fit for $T \leq 5$~K yields $\theta_{\text{CW}} = -0.2(1)$ K. The inset shows the entire inverse susceptibility curve from 2-300~K along with the simulated crystal field contribution as discussed in the text. Error bars are contained within the plotted linewidth.}
        \label{fig:Susceptibility}
\end{figure}

Susceptibility measurements are presented in Fig.~\ref{fig:Susceptibility} with data taken for 2 $<$ $T$ $<$ 5~K in the main figure, and 2 $<$ $T$ $<$ 300~K in the inset figure. Measurements have been performed in an applied magnetic field of 0.1 T. The crystal field parameters are strong, and consequently only the ground state doublet is occupied at the lowest temperatures, $T \leq 5$~K. In fact, the susceptibility for $T \geq 5$~K is well reproduced by crystal field calculations, neglecting exchange interaction. In these calculations, we use the Stevens parameters as obtained by Pearson et al. \cite{Pearson_1967}, and verified from our IN4 experiment. Data and model are shown in the inset of Fig.~\ref{fig:Susceptibility}, for more details, see appendix \ref{appendix_CF}. 

The effects of the exchange interaction on the susceptibility become prominent for temperatures below 5~K, when the crystal field levels no longer dominate. Figure \ref{fig:Susceptibility} shows a linear fit to the inverse magnetic susceptibility for $T \leq 5$~K. A Curie Weiss temperature $\theta_{\text{CW}} = -0.2(1)$~K, is extracted, indicative of weak AFM interactions. This result is in contrast to the FM interactions determined by Filippi et al.\ \cite{Filippi_1980}.
 
\subsection{Neutron scattering}\label{sec:results:scattering}
\subsubsection{Thermal neutron spectroscopy}\label{test}
\begin{figure}[!h]
        \includegraphics[width = \linewidth]{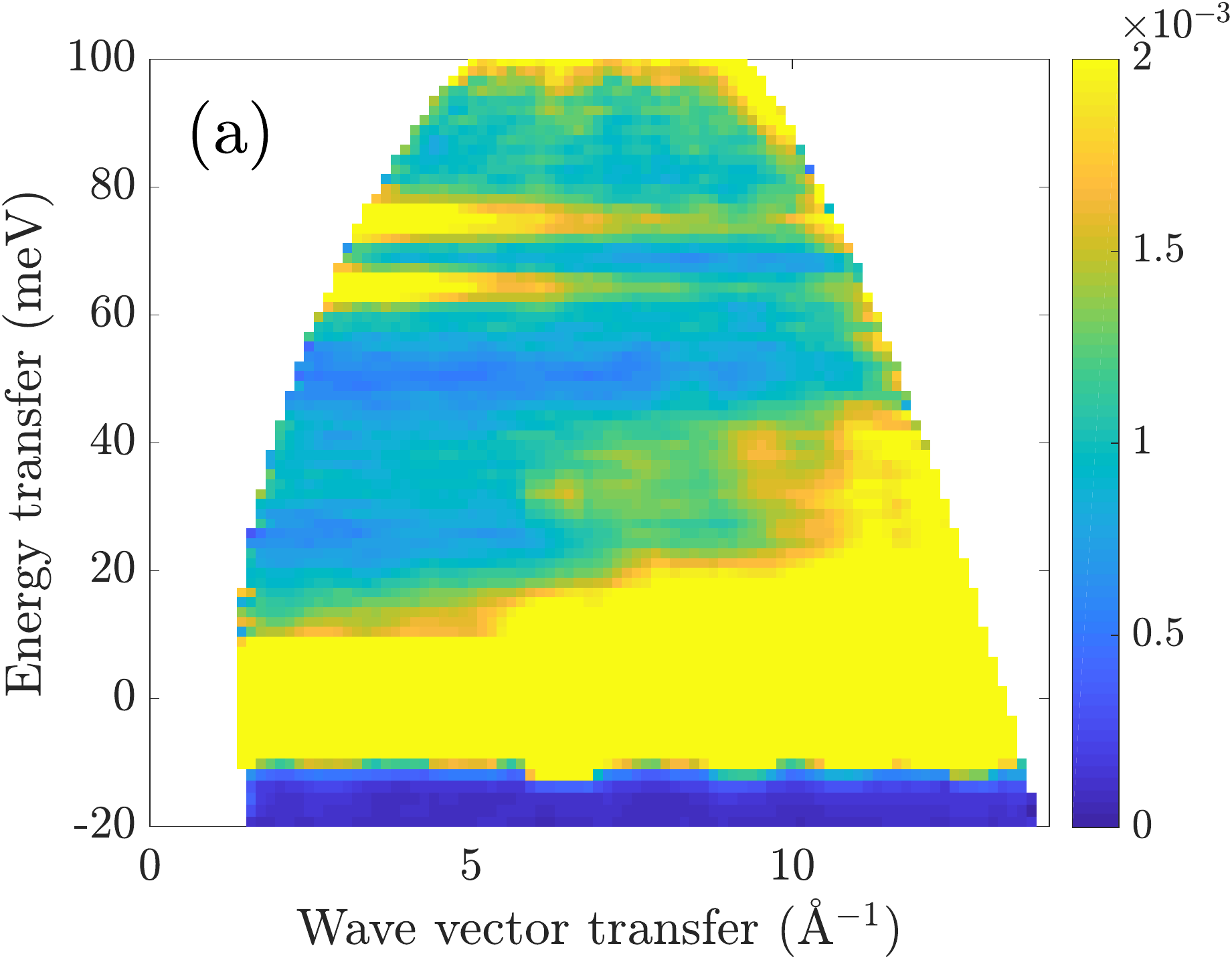}
                \includegraphics[width = \linewidth]{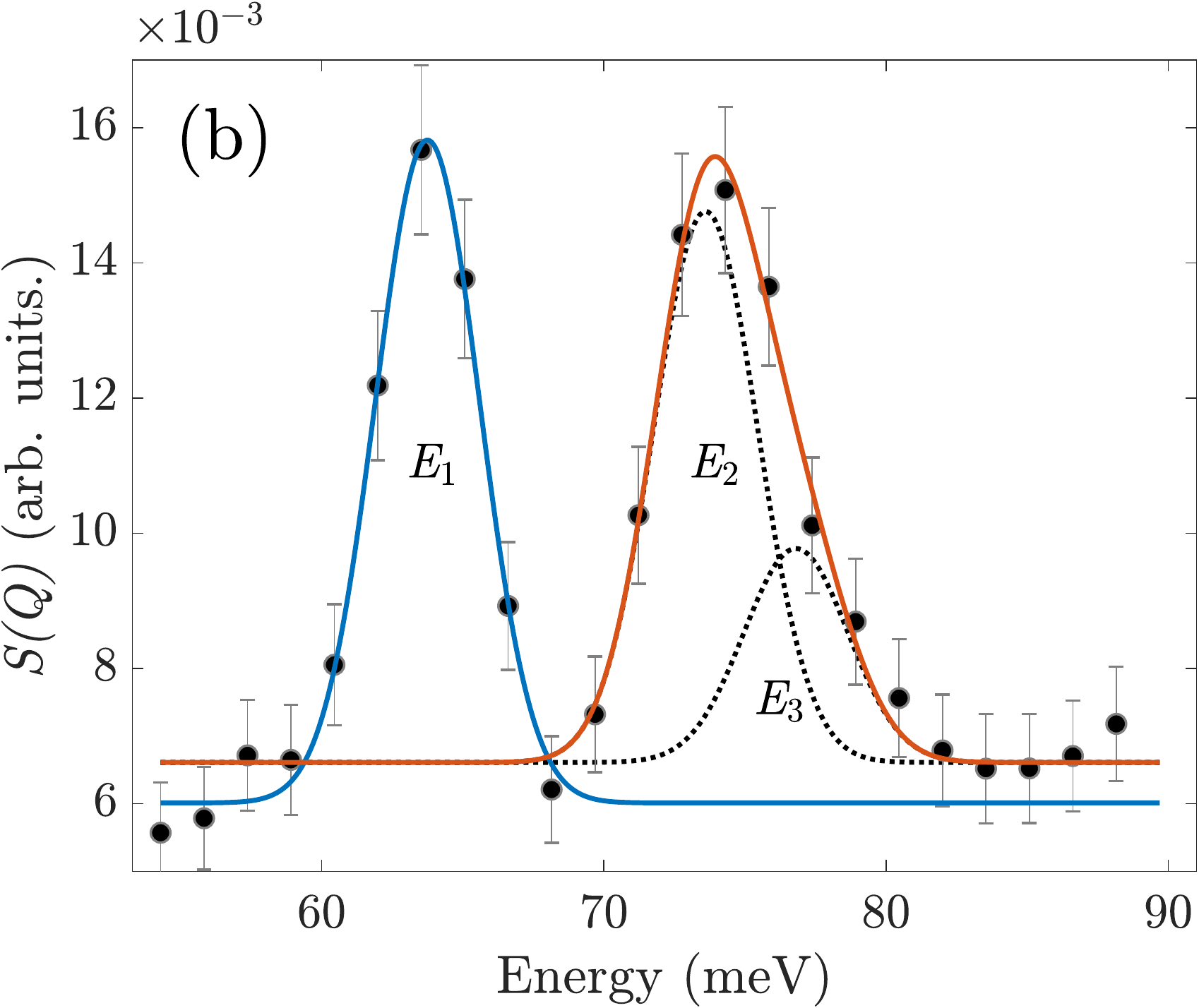}
        \caption{ $S({\textbf{Q}, \omega})$  of the crystal field excitations in YbGG showing the excitations well separated from the ground state doublet.  Colorbar represents neutron scattering intensity.(b) Integrated data for 4 $\leq$ Q $\leq$ 5 $\AA^{-1}$. The two upper excitations (E2,E3) are resolved using a double Gaussian lineshape.}
    \label{fig:IN4_2D}
\end{figure}

In YbGG the Yb$^{3+}$ ion is surrounded by eight nearest neighbour oxygen ions and therefore experiences a dodecahedral local environment and an orthorhombic site point symmetry. The relevant crystal field levels in YbGG can be most accurately determined via inelastic neutron scattering. Figure~\ref{fig:IN4_2D} presents inelastic neutron scattering data with an incident neutron energy $E_{\rm i} = 113$~meV. As expected, three crystal field excitations are located at energies $E_1 = 63.8(2)$~meV, $E_2 = 74(1)$~meV and $E_3 = 77(2)$~meV, respectively. The two upper excitations are not fully resolved, with the highest excitation appearing as a shoulder on the second excitation. All three excitations are dispersionless and follow the Yb$^{+3}$ form factor, expected for the single ion effect of a crystal field excitation. The excitation energies closely match previous experimental\cite{Buchanan_1967} and theoretical\cite{Pearson_1967} results. Based on these results we confirm the isolated $\Gamma_7$ doublet ground state of the Yb$^{3+}$ spins in YbGG with corresponding g-factors gx=2.84, gy=3.59 and gz=-3.72. YbGG is therefore an effective spin $S=1/2$ system at low temperatures $T \leq 5$~K. The crystal field analysis is further described in appendix~\ref{appendix_CF}. 

\subsubsection{Cold neutron spectroscopy}
\begin{figure}[htp!!]
	\centering
       \includegraphics[width = \linewidth]{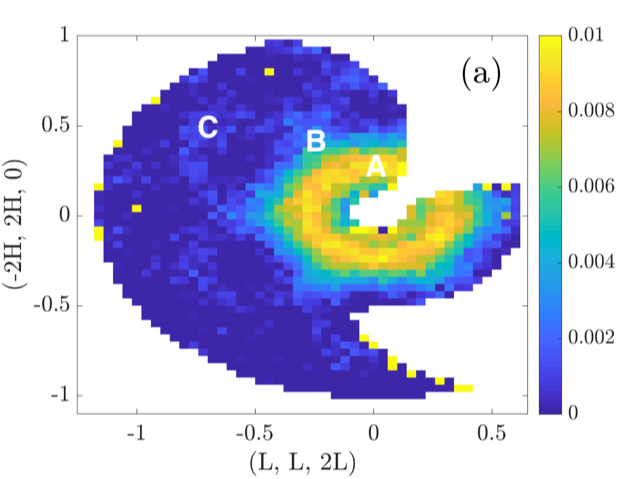}
    \includegraphics[width = \linewidth]{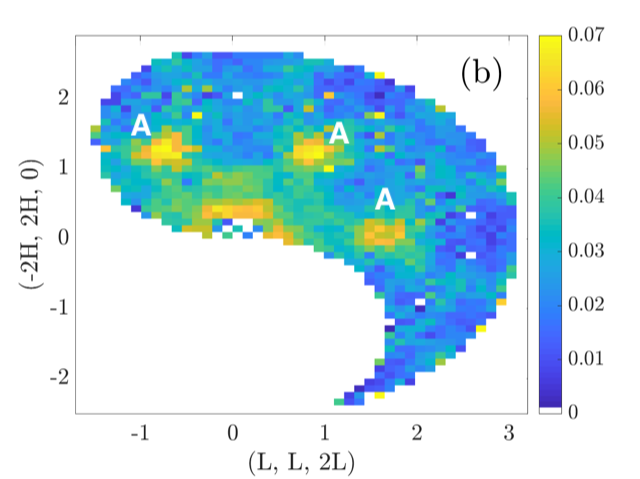}
    \includegraphics[width = \linewidth]{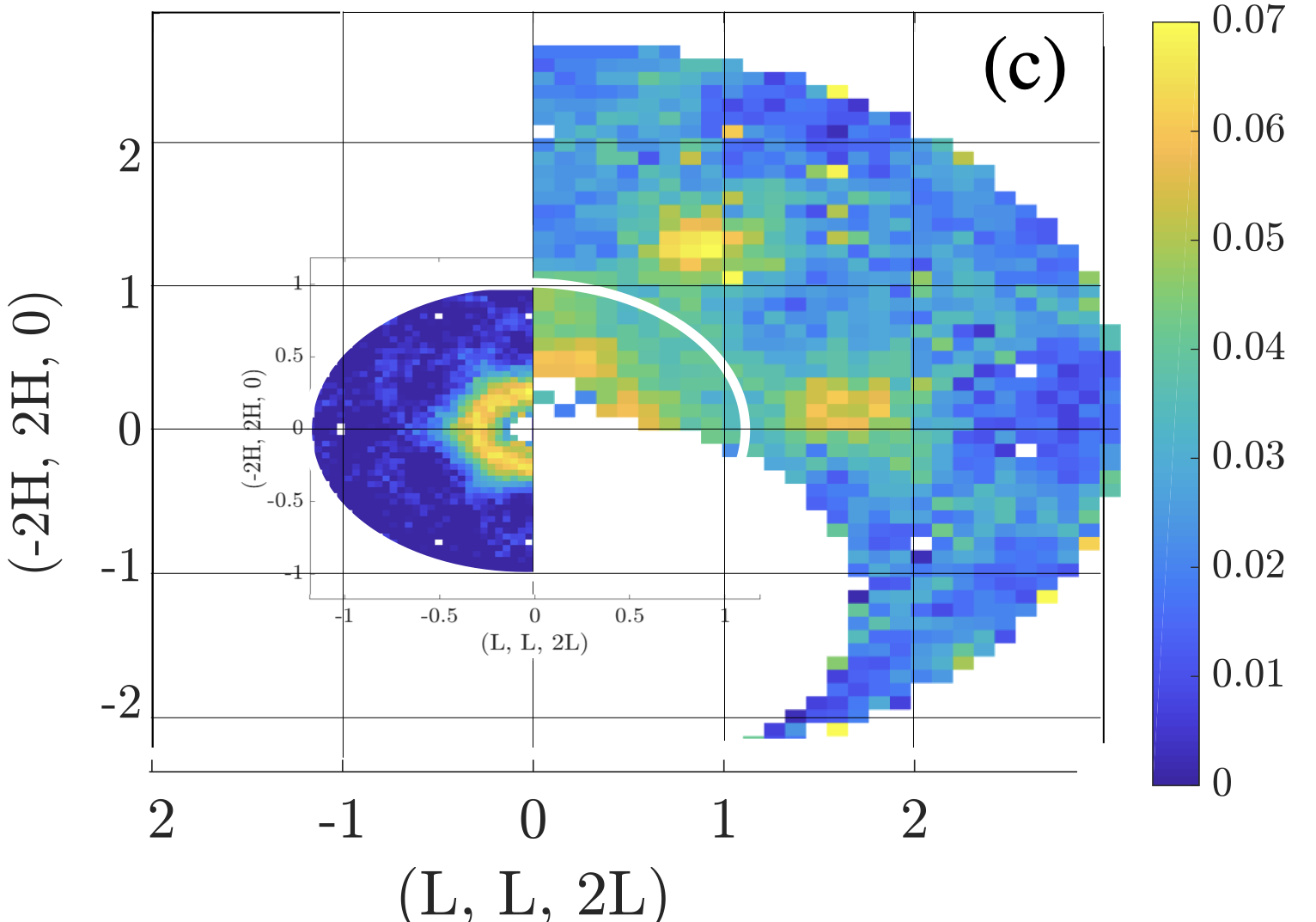}
	\caption{(a) $S_{\rm magff}(\textbf{Q})$, $E_{\rm i} = 1.55$~meV derived from a high temperature subtraction. (b)$S_{\rm mag}(\textbf{Q})$, $E_{\rm i} = 8.11$~meV. We estimate the sample temperature to be $0.1 < T < 0.2$~K. (c) Relative regions of reciprocal space probed in $S_{\rm magff}(\textbf{Q})$, $E_{\rm i} = 1.55$~meV (CNCS) (a) and $S_{\rm mag}(\textbf{Q})$, $E_{\rm i} = 8.11$~meV (D7) (b).}
    \label{fig:diffract}
\end{figure}

The magnetic energy scales in YbGG are in the mK regime and thus accessible via cold neutron scattering.  Figure~\ref{fig:diffract}(a) presents the magnetic contribution to the elastic scattering profile measured at CNCS with incoming neutron energy $E_{\rm i} = 1.55$~meV, accessing a low Q region. The elastic magnetic scattering profile, $S_{\rm magff}(\textbf{Q})$, is extracted from the scattering within the instrumental energy resolution with a background subtraction of equivalent scattering at 13 K, in the paramagnetic regime. In comparison, Fig.~\ref{fig:diffract}(b) presents the magnetic contribution measured on D7, $S_{\rm mag}(\textbf{Q})$, of the energy integrated measurements with $E_{\rm i} = 8.11$~meV and thus provides a wider $\textbf{Q}$ range. The magnetic signal is extracted using XYZ polarisation analysis \cite{D7_Overview} from the spin flip channel. $S_{\rm magff}(\textbf{Q})$ can therefore be considered as a static contribution. Fig.~\ref{fig:diffract}(c) shows the relative regions of reciprocal space accessed by the CNCS and D7 datasets and their overlap. The Q range and Q resolution accessed in the experiments vary significantly due to the different incident wavelength.  The CNCS data set, figure~\ref{fig:diffract}(a), extends across 0.1 $<$  (2H, 2H, 0) $\sim$ and 0.13 $<$ (L,L,2L) $\sim$ 1. In contrast, the D7 data set extends across 0.3 $<$ (2H, 2H, 0) $<$ 3 and 0.2 $<$ (L,L,2L) $<$ 3. Of course, the relative $\textbf{Q}$ resolutions also vary significantly affecting boundary conditions and smoothing features in the D7 data that are distinct in the CNCS data.  Both data sets show distinct, non resolution limited, short range correlated scattering, the \textbf{Q} dependence of which does not follow the magnetic form factor of Yb$^{3+}$. Indeed the scattering is correlated with a 6-fold symmetry, consistent with the crystalline structure. The short ranged nature of the magnetic structure factors measured, consistent with the broad feature in the specific heat data, provides confidence that a sample temperature of 0.1 < T < 0.6 K was reached. 
 Figure~\ref{fig:diffract}(a), with the highest Q resolution, shows clearly a hexagon feature for $\vert Q \vert \leq 0.63$~\AA$^{-1}$, marked $\textbf{A}$. The reduced intensity for the lowest \textbf{Q}, $\vert Q  \vert \rightarrow 0 $~\AA$^{-1}$, indicates that these correlations are AFM. A fit to the data with a simple Gaussian lineshape, see appendix~\ref{app:SqData}, shows a peak in intensity at $\vert Q \vert = 0.30(3)$~\AA$^{-1}$ corresponding to a lattice spacing $d=2 \pi / Q = 20(2)$~\AA , and a correlation length of $12(2)$~\AA, as determined from the peak full width at half maximum (FWHM). Weak circular features extend from the edges of the hexagon, $\textbf{B} \rightarrow \textbf{C}$. The low \textbf{Q} hexagon feature is also visible in the D7 data but is limited due to reduced \textbf{Q} resolution and detection boundaries. The higher \textbf{Q} D7 data shows three distinct diffuse peaks, figure~\ref{fig:diffract}(b) $\textbf{A}$, centered at $\vert Q \vert = 1.95(8)$~\AA$^{-1}$, corresponding to a lattice spacing of $d= 3.2(1)$~\AA , with a correlation length of $4(0.6)$~\AA. These peaks,$\textbf{A}$, also follow the 6-fold symmetry of the crystalline structure. In a simplistic manner, considering the Q positions and correlation lengths, one could assign the low Q hexagonal feature to a looped structure encompassing ten ions extending throughout the unit cell while the higher Q features are derived from near neighbor exchange.

\section{Data Modelling}\label{sec:analysis}
\subsection{Reverse Monte Carlo}\label{sec:analysis:RMC}

\begin{figure}[h!]
	\centering
      \includegraphics[width =  \linewidth]{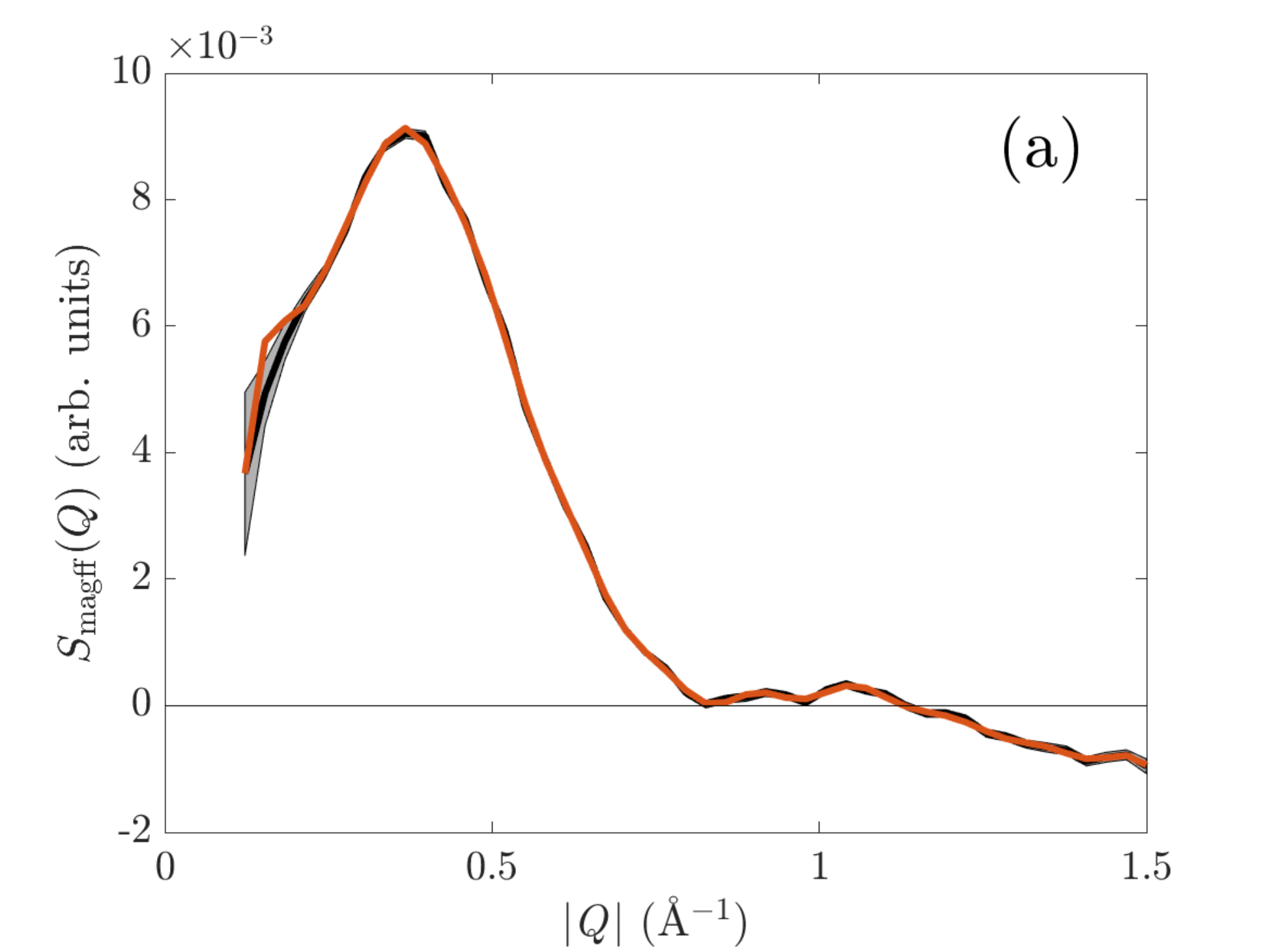}
    \includegraphics[width = \linewidth]{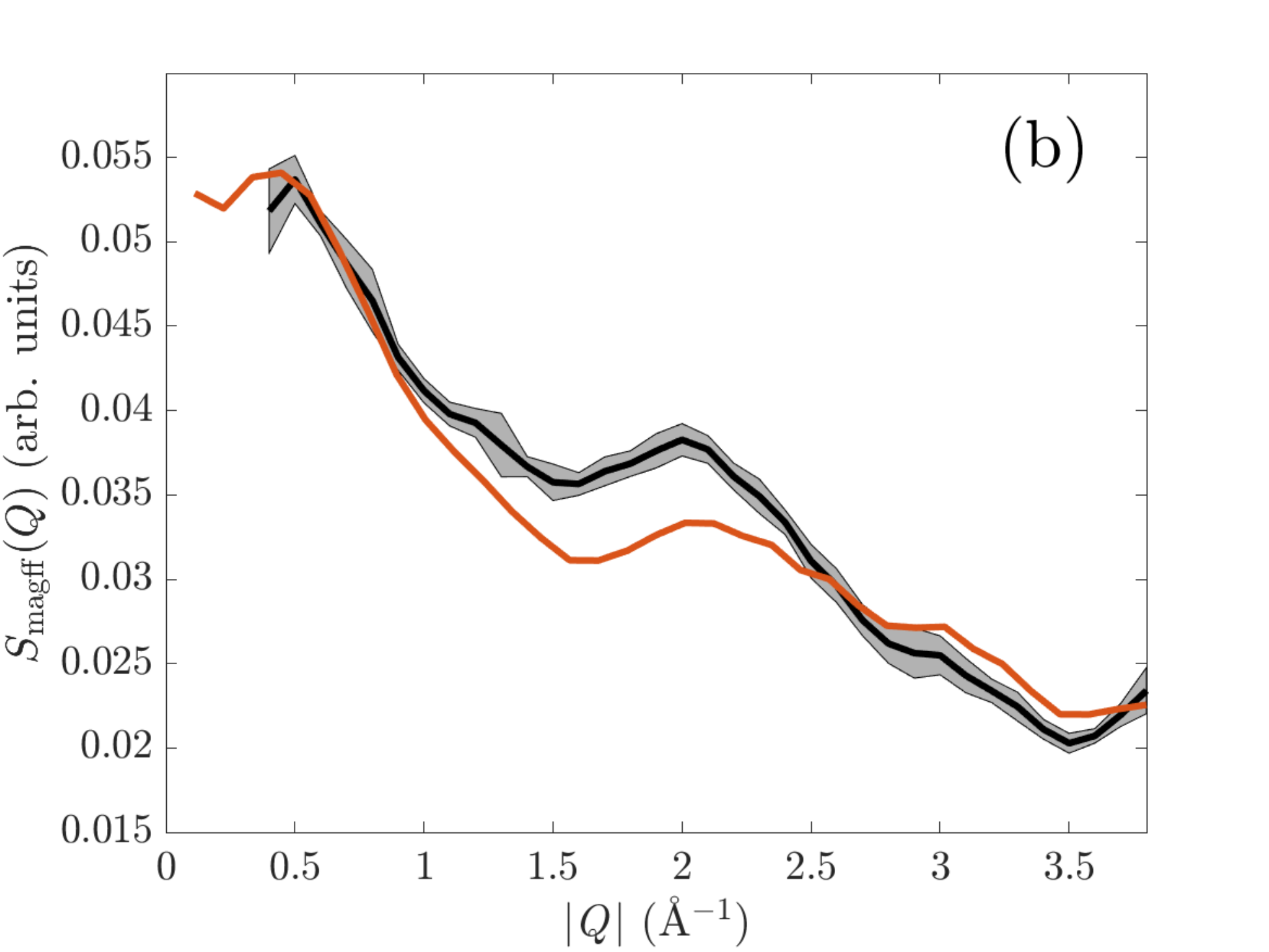}
	\caption{Comparison of $S({Q})$ RMC simulation (red) and powder averaged data (black). (a) Powder averaged CNCS data and RMC $S({Q})$ simulation. (b) Powder averaged D7 data and RMC $S({Q})$ simulation. }
    \label{fig:spinvert:1D}
\end{figure}

We have performed RMC simulations on $S_{\rm magff}(\textbf{Q})$ and $S_{\rm mag}(\textbf{Q})$.  It is, however, not possible to directly minimise the 2D $S(\textbf{Q})$ of the single crystal results, since the RMC simulations leaves all points in \textbf{Q} outside the (-2H, 2H, 0), (L, L, 2L) scattering plane unconstrained and can thus lead to errors. In this work, we have mitigated the possibility of erroneous minimisation with three approaches. First, comparing data from several experiments with various incident energies and thus energy and \textbf{Q} resolution. Second, creating an isotropic scattering distribution from the measured 2D $S(\textbf{Q})$ through integration of all points with similar $\vert Q \vert$, which we shall term powder diffraction pattern $S({Q})$, see Fig.~\ref{fig:spinvert:1D}, and deriving a single crystal pattern, $S(\textbf{Q})$, from the RMC spin configuration obtained. Third, we use the average of 400 RMC minimisations to obtain good statistics on the spin correlations. We accept that the data presented is only an approximation of the true correlations. By testing several extrapolation techniques in addition to extracting RMC from various data sets with different incident energies and averaging across 400 RMC minimizations, we believe that the results are stable and that some variation in the assumed extrapolation will not affect the fundamental structure of the solution. The exact procedure is outlined in appendix \ref{app:RMC}. 

Figure \ref{fig:spinvert:1D} compares the result of the RMC simulation with the $S({Q})$ powder diffraction pattern from CNCS, $S_{\rm magff}(Q)$, (a) and D7, $S_{\rm mag}({Q})$, (b). Figure \ref{fig:spinvert:1D}(a) shows an excellent reproduction of powder data for all $Q$. 
In contrast, the reproduction of the D7 powder diffraction pattern in Fig.~\ref{fig:spinvert:1D}(b) provides reasonable agreement only for $Q \leq 0.8$~\AA$^{-1}$. For higher $Q$, the RMC model shows similar features, but with discrepancies in the intensities. We do not simultaneously minimize the CNCS and D7 datasets since this would introduce an additional parameter representing the importance of each dataset and the various regions of reciprocal space. Our approach is minimalistic and shows the extreme cases when minimising to the respective data sets. 
\begin{figure}[h!]
	\centering
    \includegraphics[width = 0.9\linewidth]{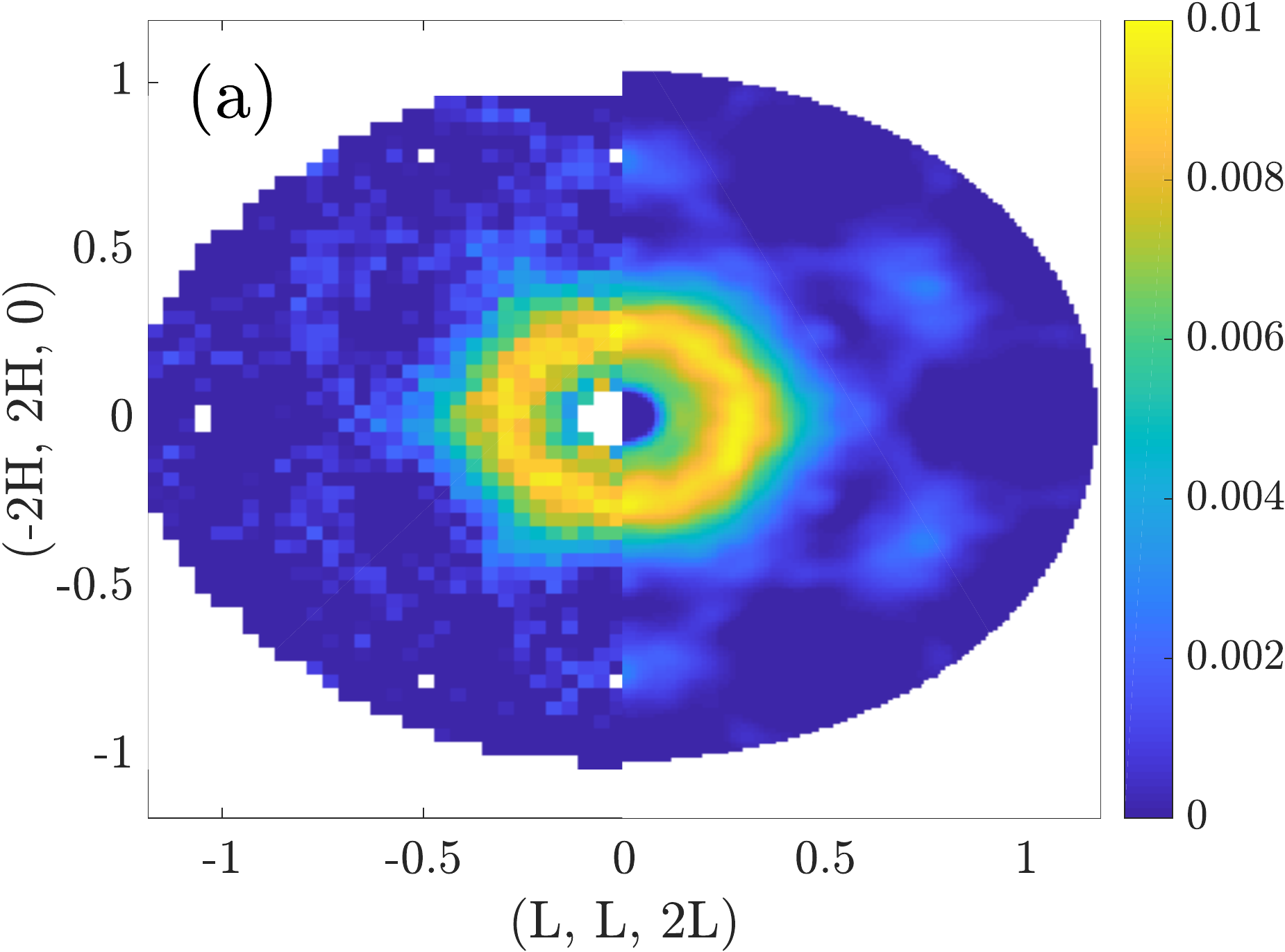}
    \includegraphics[width = 0.9 \linewidth]{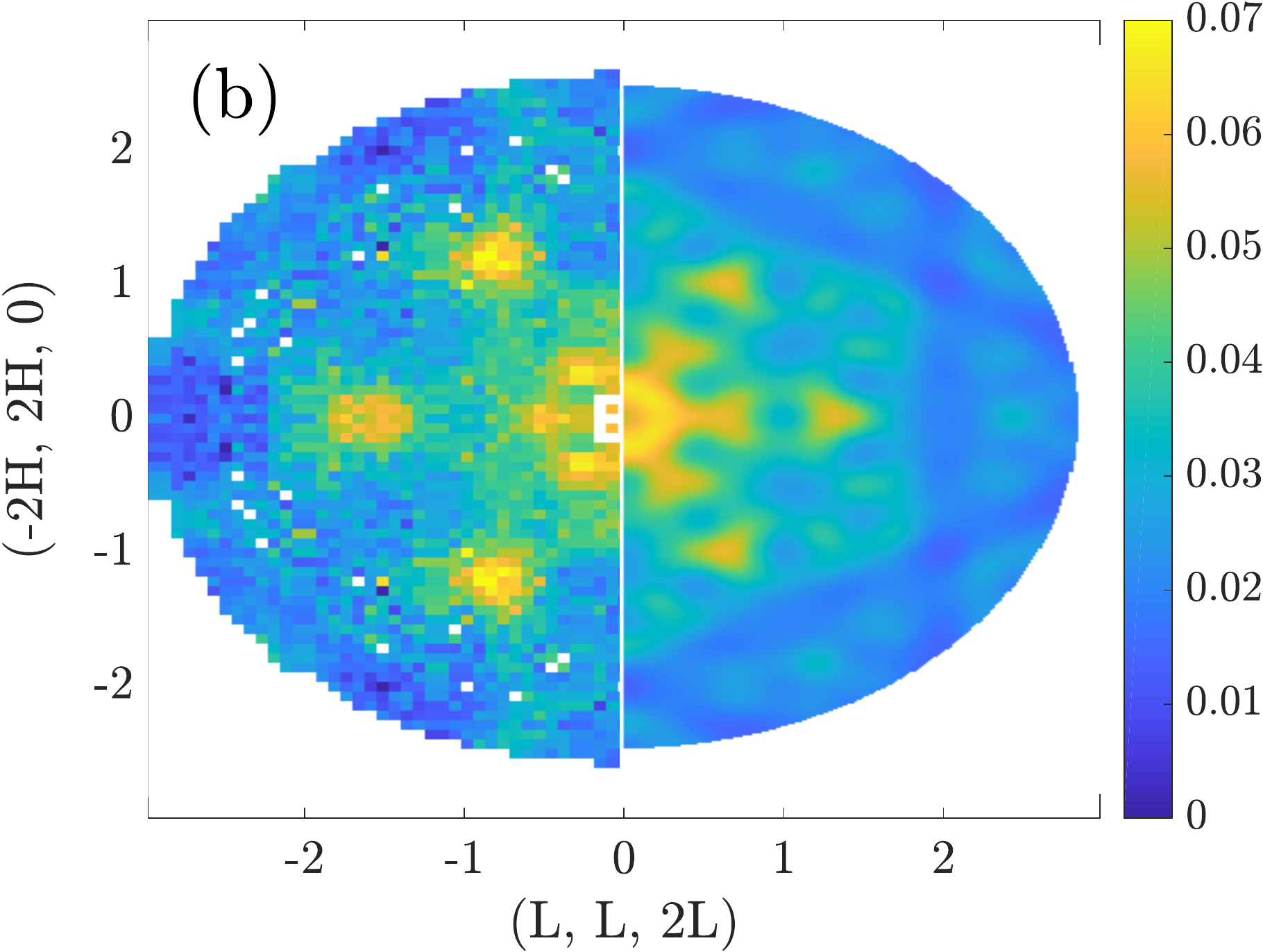}
	\caption{Comparison of experimental data and RMC fit (left and right respectively). (a)CNCS  $S_{\rm magff}(\textbf{Q})$ (b) D7 $S_{\rm mag}(\textbf{Q})$. Colorbar represents scattering intensity.}
    \label{fig:spinvert}
\end{figure}

The spin structure derived from the RMC $S({Q})$ powder refinement is used to recalculate the 2D magnetic scattering profiles, $S_{\rm magff}(\textbf{Q})$ or $S_{\rm mag}(\textbf{Q})$, and subsequently compared to experimental data, see Fig.~\ref{fig:spinvert} for CNCS $S_{\rm magff}(\textbf{Q})$ data (a) and D7 $S_{\rm mag}(\textbf{Q})$ data (b).
The RMC $S_{\rm magff}(\textbf{Q})$ of the CNCS data contains the correct crystal symmetry and accurately reproduces all of the main features at the correct \textbf{Q} positions, including the low-$Q$ hexagon and the higher-$Q$ features extending from the sides of the hexagon. In contrast, the comparison in Fig.~\ref{fig:spinvert}(b), of the D7 data and the corresponding RMC $S_{\rm mag}(\textbf{Q})$, is much less accurate. Although the main features are reproduced, the broad $Q$ features are slightly offset. 

There are several subtle differences between the CNCS and D7 neutron scattering intensities that may give rise to the difference in accuracies. The CNCS magnetic scattering intensity, $S_{\rm magff}(\textbf{Q})$, is obtained via the subtraction of a high temperature scattering from base temperature scattering. The high temperature scattering provides an intense magnetic formfactor and $S_{\rm magff}(\textbf{Q})$ can result in negative intensities. This is considered in the RMC. D7 magnetic scattering, $S_{\rm mag}(\textbf{Q})$, is extracted using XYZ polarisation. The determination of $S_{\rm mag}(\textbf{Q})$ in this manner assumes that the net moment of the compound is zero as is the case for paramagnetic systems or powdered antiferromagnet compounds. A ferromagnetic signal would induce some depolarisation of the scattered polarization. Using this equation for the case of a single crystal makes an implicit assumption that there is a net zero averaged moment with no symmetry breaking such that the magnetic cross-section is isotropic with magnetic components of equal magnitude projected along the three orthogonal directions. We made these assumption since (a) we did not observe any depolarisation of the scattered beam, (b) only short range order was observed and (c) prior knowledge of the director state which provides an isotropic spin distribution, to a first approximation. However, the incoherent scattering signal, expected to be homogeneous in $\textbf{Q}$, contains weak hexagonal features reminiscent of the magnetic signal that affect only the peak intensities of $S_{\rm mag}(\textbf{Q})$. RMC optimises directly to S(Q) and is sensitive to such relative changes. We suggest that these small variations give rise to the differences observed between the CNCS and D7 RMC and is the origin of the poorer simulations of the D7 data. Nevertheless, the resultant D7 RMC spin structure is consistent with that determined from the CNCS RMC and provides confidence in our results.

In order to interpret the RMC results, the spin distributions and correlations are investigated. In the following, only CNCS RMC simulations are presented, but despite the less perfect correspondence between RMC results and D7 data, there is strong equivalence between the spin distributions and correlations obtained from the RMC derived spin structure of all datasets, see appendix \ref{app:RMC}. The resultant D7 RMC spin structure, is consistent with that determined from the CNCS RMC, and distinctly different from the spin structure determined for GGG, see Fig. \ref{Richard:fig:phiDistribution}. The similarities between the spin structures extracted from different experiments with very different Q ranges and resolutions provides confidence in our results. 

\newpage
\begin{figure}[h!]
	\centering
      \includegraphics[width = 0.9\linewidth]{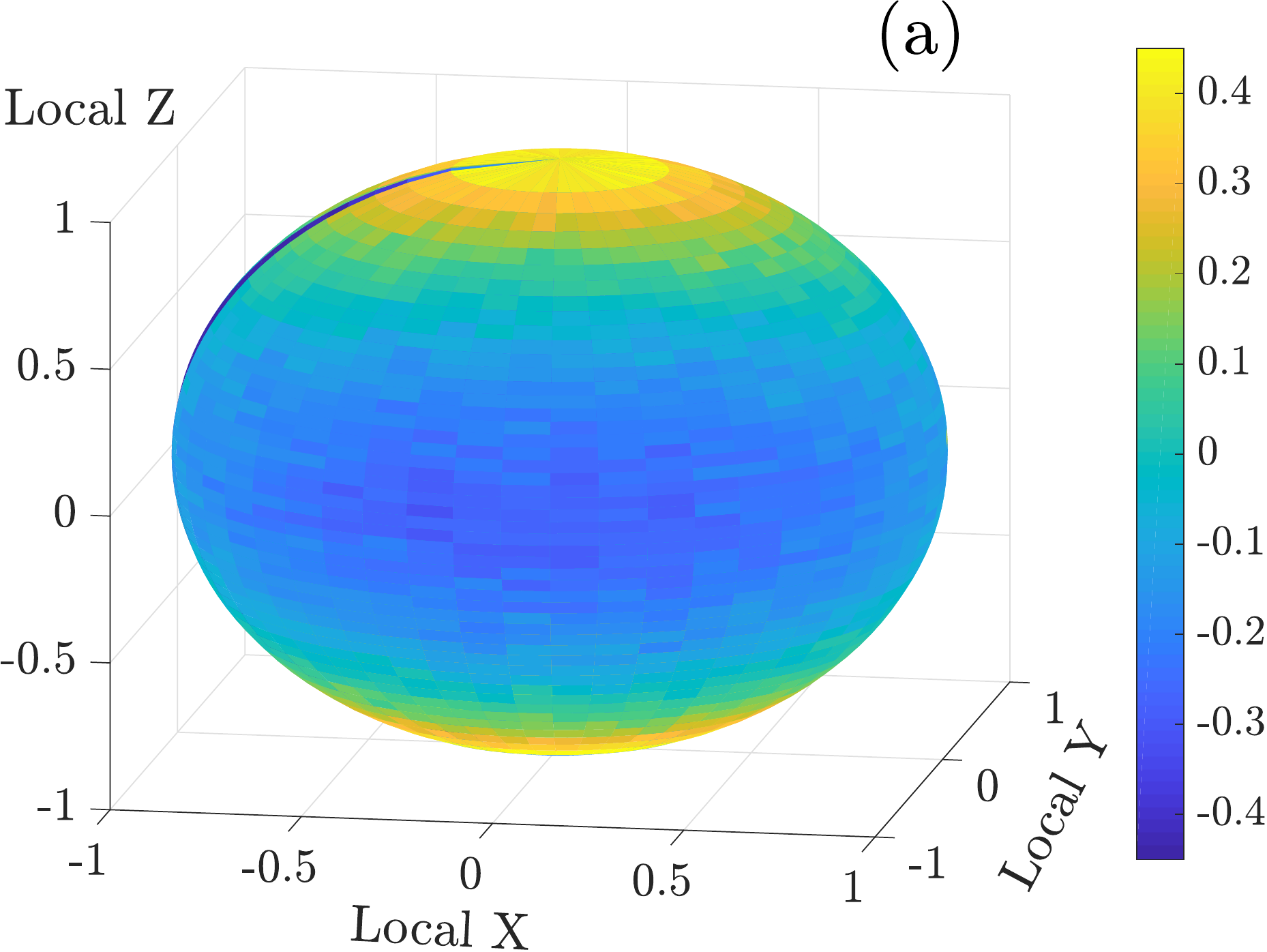}
    \includegraphics[width = \linewidth]{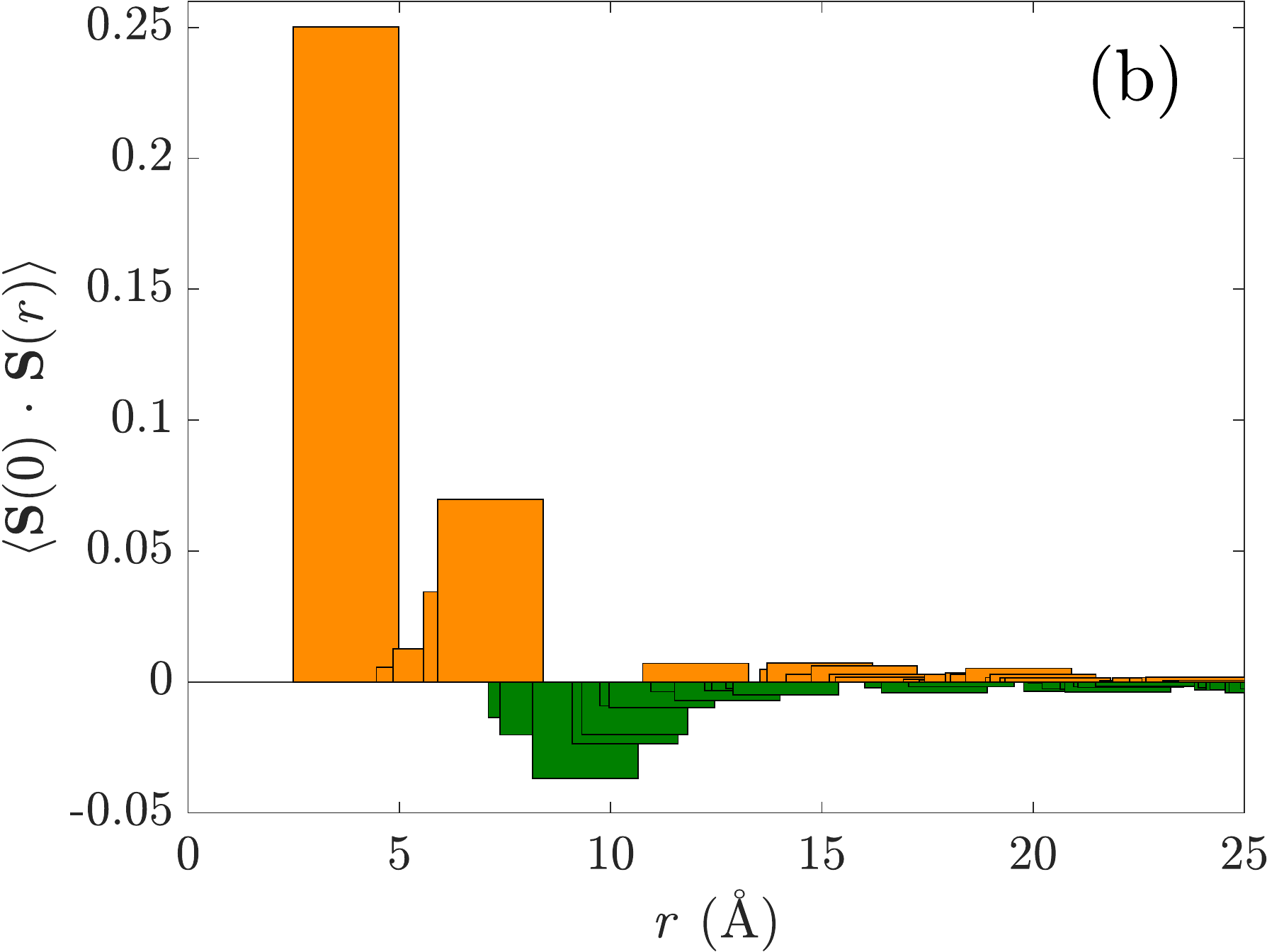}
	\caption{(a) Stereographic projection of the spin distribution in the local coordinate system with a log color scale. The spins show an easy axis along the local $z$-direction. (b) Radial dependence of $\langle \textbf{S}(0)\cdot \textbf{S}(r) \rangle$. Positive scalars orange, negative scalars green.}
    \label{fig:SiSj}
\end{figure}
\begin{figure}[h!]
	\centering
    \includegraphics[width = \linewidth]{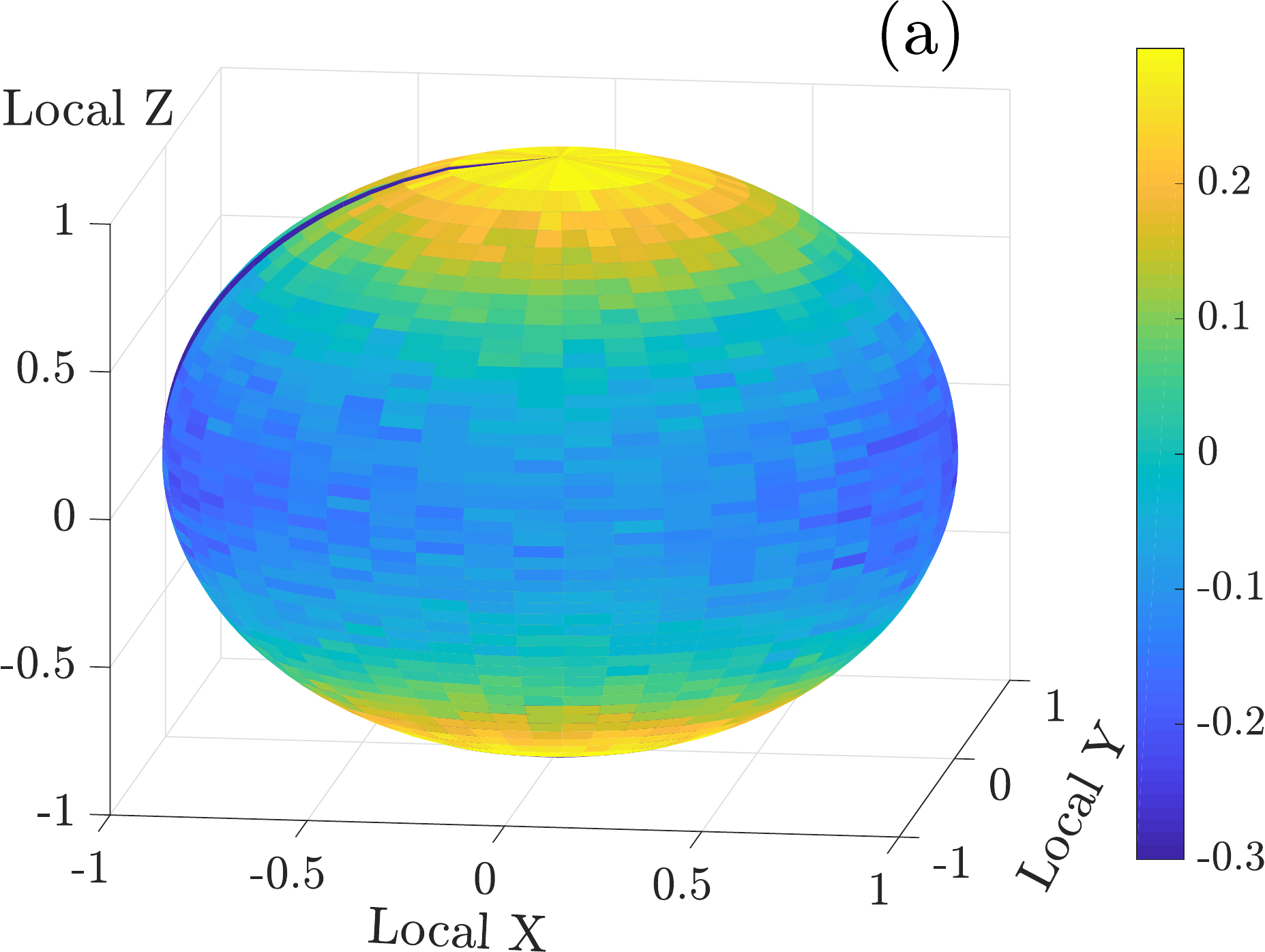}
   \includegraphics[width =\linewidth]{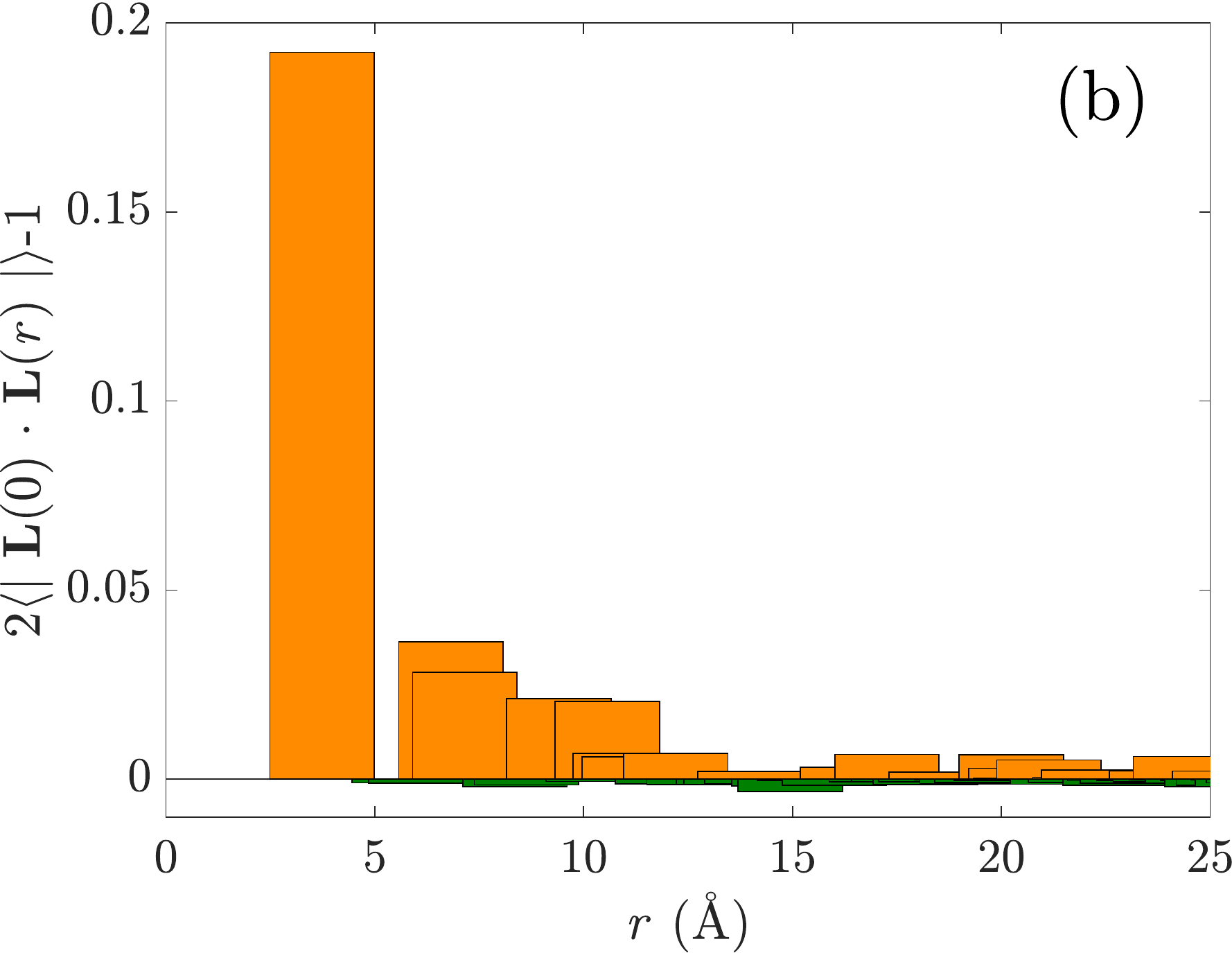}  
   	\caption{(a) Stereographic projection of the director distribution in the local coordinate system with a log colour scale. (b) Radial correlation function of the directors. Positive scalars are plotted in orange, negative scalars are plotted in green.}
        \label{fig:LiLj}
\end{figure}

Figure \ref{fig:SiSj}(a) presents the spin probability distribution, derived from RMC, in the local coordinate system showing an easy axis along the local $z$-direction, the axis which connects the centers of two adjacent triangles within the crystal structure, see Fig.~\ref{fig:YbGG}(right). Figure \ref{fig:SiSj}(b) presents the average spin-spin correlations $\langle \textbf{S}(0)\cdot \textbf{S}(r) \rangle$ as a function of spin-spin distance. NN correlations are on average positive and thus FM with an average angle of $73^{\circ}$ between neighbouring spins. This is in contrast to the AFM NN correlations and strong planar anisotropy in the local XY plane observed for GGG \cite{Paddison_2015_GGG}.  Figure \ref{fig:SiSj}(b) further shows that spins in YbGG are correlated AFM across the loop, consistent with FM NN correlations and corresponds to the spatial scales extracted from the low Q hexagon, Fig.~\ref{fig:diffract}(a). This final spin structure results in a director state. We find that the local easy axis of the directors are along the local $z$-direction, see Fig.~\ref{fig:LiLj}(a), directly equivalent to the director state found in GGG. The director state is further supported by the magnetic excitations observed in the extended CNCS data set, see Fig.~\ref{fig:sqw}. Three dispersionless low lying excitations are observed at 0.06, 0.1 and 0.7 meV entirely consistent with dispersionless excitations observed in GGG and assigned to the director state \cite{Deen2010,Ambrumenil2015}. Detailed analysis of the excitation spectra will be published elsewhere. 

We next investigate the correlations between the directors, $\textbf{L}$, Eq.~(\ref{eq:director}). The radial correlation function of the directors $g_L=2\langle\hat{\textbf{L}}(0)\cdot \hat{\textbf{L}}(r)\rangle-1$ is equal to $-1$ if, on average, the loop directors are orthogonal to each other and to $+1$ if collinear. Figure \ref{fig:LiLj}(b) shows the radial correlation function and reveals a predominantly collinear director state within the first unit cell, 12.2 \AA. However, unlike the long range correlated state of GGG, the directors in YbGG correlate weakly beyond the first unit cell. 

\begin{figure}[h!]
	\centering
     \includegraphics[width = \linewidth]{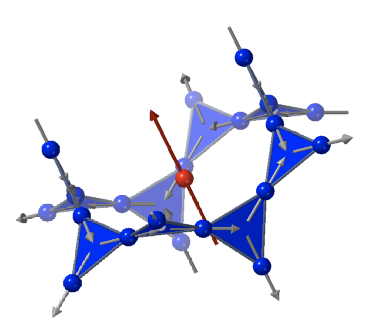}
	\caption{A 10-spin-loop together with a single ion from the opposite hyperkagome lattice (central, red). The blue spheres depict 
	Yb$^{3+}$ ions while the red sphere can be considered as the net average magnetic moment of the ten-ion loop, the director. Local spin distributions peak along the local $z$-direction (grey arrows) which connects the centers of adjacent triangles. The local spin structure is presented with spins point along the easy axis. The director distribution (red arrow) peaks along the local $z$-direction. }
    \label{fig:Concept}
\end{figure}

The resultant spin configuration and director state in YbGG are presented in Fig.~\ref{fig:Concept} which shows FM correlated NN spins along the local easy axes as well as the resultant director of the loop.


\subsection{Monte Carlo}\label{sec:analysis:MC}
\begin{figure}[!h]
	\centering
	\includegraphics[width = \linewidth]{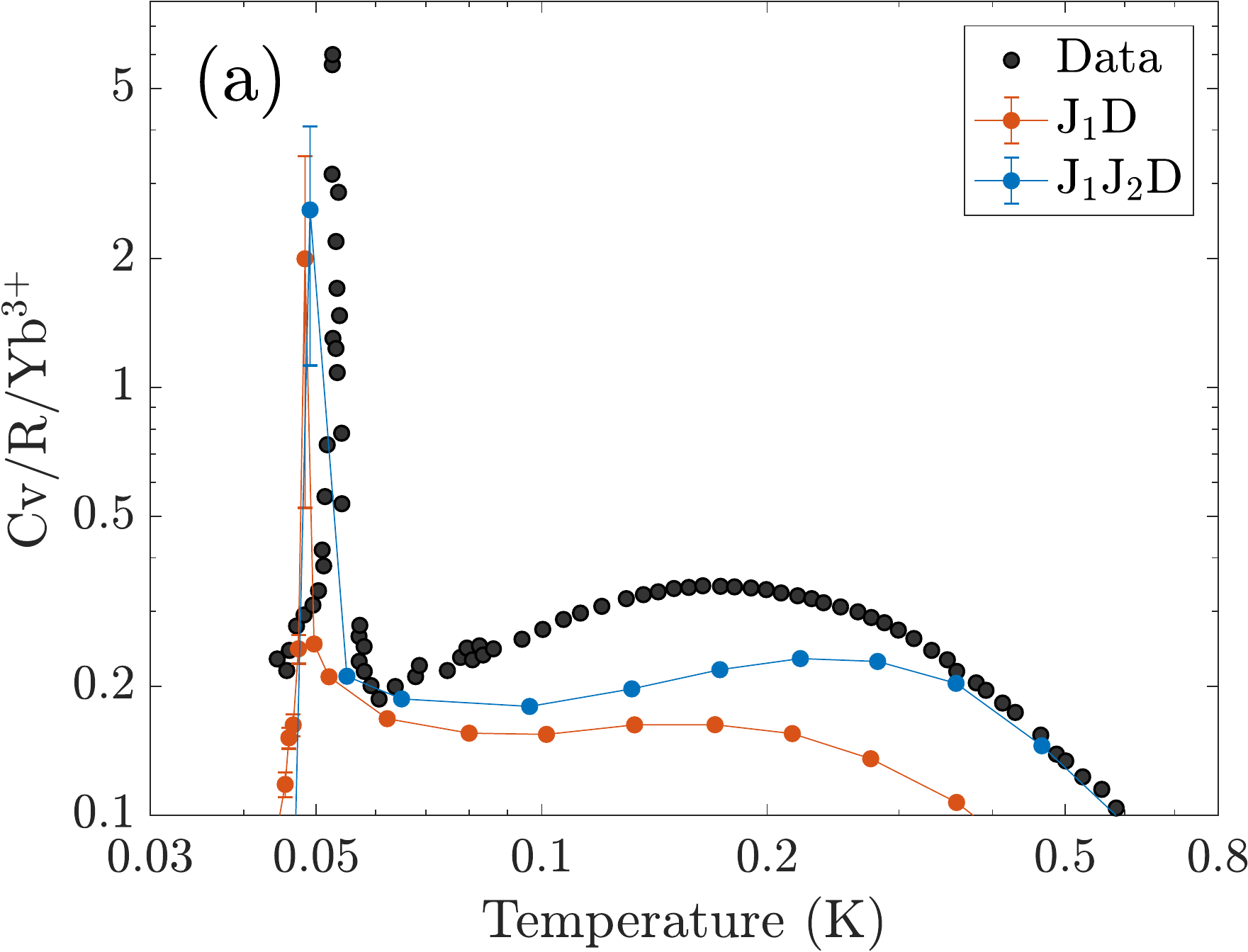}
    \includegraphics[width = \linewidth]{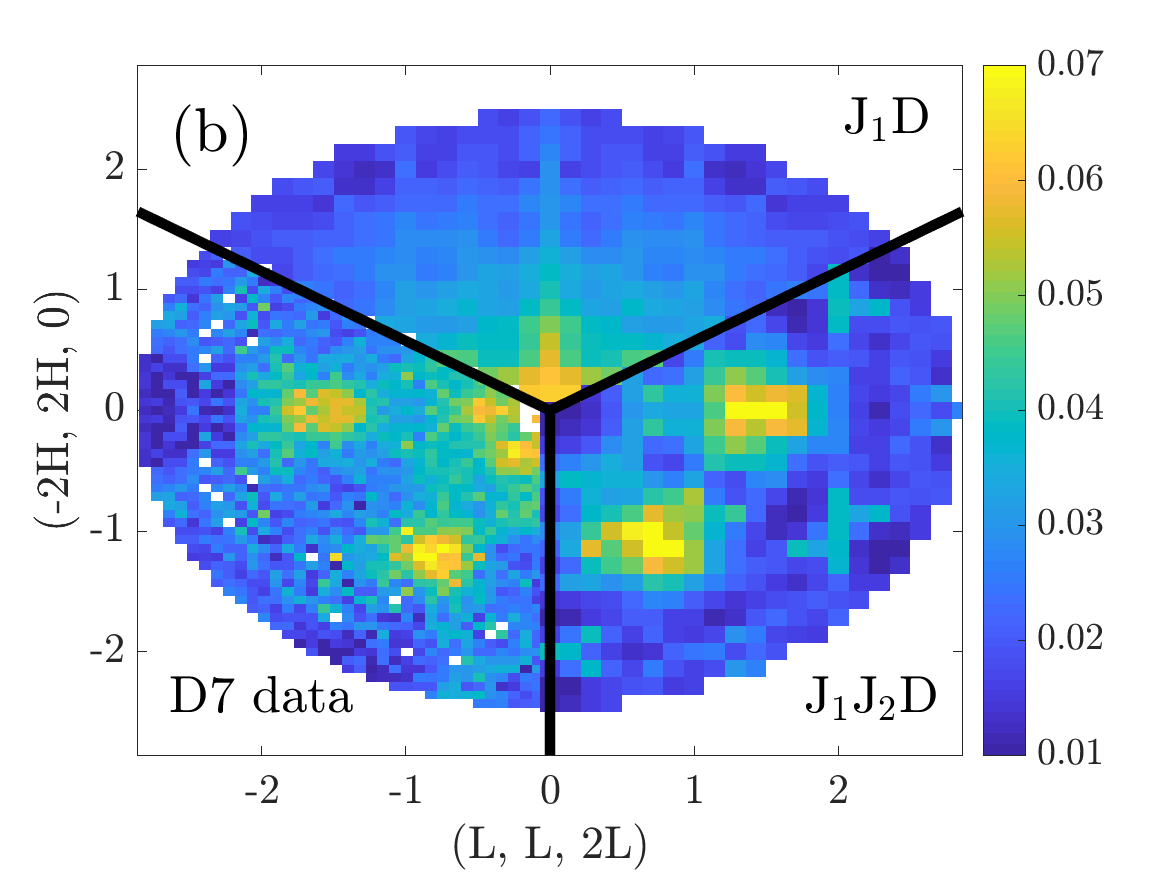}
	\caption{(a) heat capacity data\cite{Filippi_1980} with simulated heat capacity for the  $J_1D$ and  $J_1J_2D$ models. (b) Simulated $S(\textbf{Q})$, $T = 0.2$~K, for the $J_1D$ model and the $J_1J_2D$ model with $S_{\rm mag}(\textbf{Q})$. Colorbar represents $S(\textbf{Q})$.}
    \label{fig:MC:C:Sq}
\end{figure}

In order to gain a further grasp on the absolute energy scale of the spin-spin couplings in YbGG, we have investigated classical Heisenberg and Ising models with anisotropy along the local $z$-direction motivated by the RMC results. In appendix \ref{app:MC} we present a short discussion of an anisotropic Heisenberg model. In the current text, we present an Ising model optimised for the heat capacity measured in experiment\cite{Filippi_1980}, see Fig.\ref{fig:MC:C:Sq}(a). The resultant exchange parameters are used to recalculate S(\textbf{Q}) and these are compared to the experimental $S_{\rm mag}(\textbf{Q})$, Fig.~\ref{fig:MC:C:Sq}(b). We compare to $S_{\rm mag}(\textbf{Q})$ from D7 due to the extended $\textbf{Q}$ range provided in this dataset. In GGG the relevant Hamiltonian in the director phase includes the NN exchange $J_1$, the next nearest neighbour (NNN) exchange $J_2$ and the dipolar interaction $D$, with inter-hyperkagome coupling $J_3$  only relevant at lower temperatures to drive the spin slush state \cite{Rau_2016}. As such, the relevant Hamiltonian for YbGG in the director state is 
\begin{equation}\label{JJDHamiltonian}
    \begin{split}
		\mathcal{H}=&J_1\sum_{\langle i,j\rangle} \, \mathbf{S}_i\cdot\mathbf{S}_j+J_2\sum_{\langle\langle i,j\rangle\rangle} \textbf{S}_i\cdot\textbf{S}_j+\\
		&Da^3\sum_{i<j}\left(\frac{\mathbf{S}_i\cdot\mathbf{S}_j}{|\mathbf{r}_{ij}|^3}-3\frac{\left(\mathbf{S}_i\cdot\mathbf{r}_{ij}\right)\left(\mathbf{S}_j\cdot\mathbf{r}_{ij}\right)}{|\mathbf{r}_{ij}|^5}  \right).
	\end{split}
\end{equation}

Here, $a$ is the nearest neighbour distance, $\textbf{r}_i$ is the position of the classical Ising spin $\textbf{S}_i$ oriented along the local $z$-direction and $\textbf{r}_{ij}=\textbf{r}_i-\textbf{r}_j$. $\langle\cdot\rangle$ and $\langle\langle\cdot\rangle\rangle$ denote summation over NN and NNN respectively.

Two distinct models are simulated. First, we simulate a spin structure with $J_1$ and $D$ only, a $J_1D$ model. Second, we add $J_2$ in a $J_1J_2D$ model. In principle, the dipolar interaction strength can be calculated explicitly from the magnetic moment $\mu$ and inter-atomic distances, $D=\frac{\mu_0\mu^2}{4\pi a}=0.24\textup{ K}$ for $\mu=4.3$ Bohr magnetons\cite{YbGGFreeDipolarMoment}. However, the magnetic moment of Yb$^{3+}$ is strongly affected by the crystal field, which motivates varying the strength of the dipolar interaction in addition to the exchange interactions. The resultant magnetic moment derived within the $J_1D$ model is $\mu=4.19$ Bohr magnetons and $\mu=3.88$ Bohr magnetons for the $J_1J_2D$ model. 

 Figure \ref{fig:MC:C:Sq}(a) shows the resultant heat capacities for the two models with optimised parameters $J_1 = 0.6$~K, $D = 0.21$~K for the $J_1D$ model, and $J_1 = 0.72$~K, $J_2 = 0.12$~K, $D = 0.18$~K for the $J_1J_2D$ model. The lambda transition is well described by both models, and would correspond to long range ordering due to dipolar interactions if YbGG was an Ising system\cite{Yoshioka2004}. Both models reproduce the broad specific heat anomaly, albeit with an overall suppression. The $J_1J_2D$ model has better agreement with data above the lambda transition, and above 0.4~K the model coincides with data. 
 

\begin{figure}[!h]
	\centering
	\includegraphics[width = \linewidth]{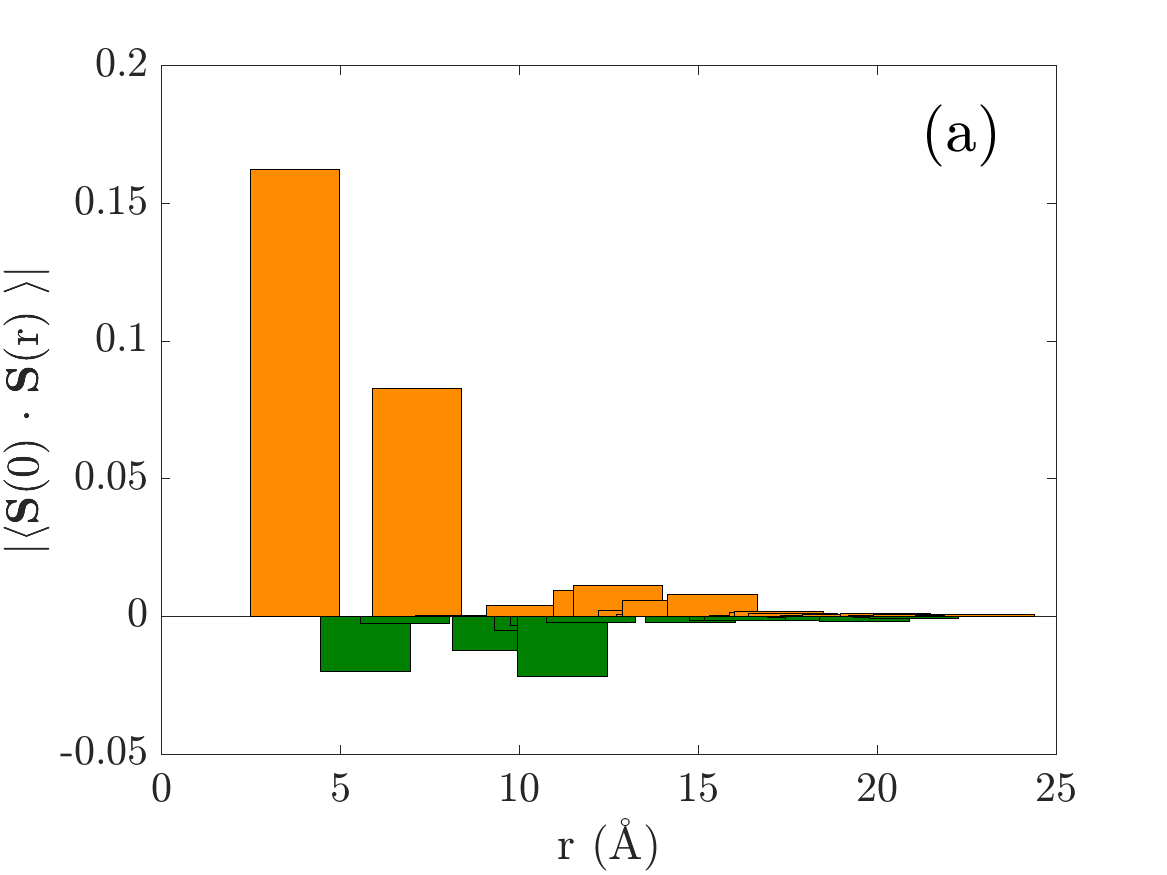}
    \includegraphics[width = \linewidth]{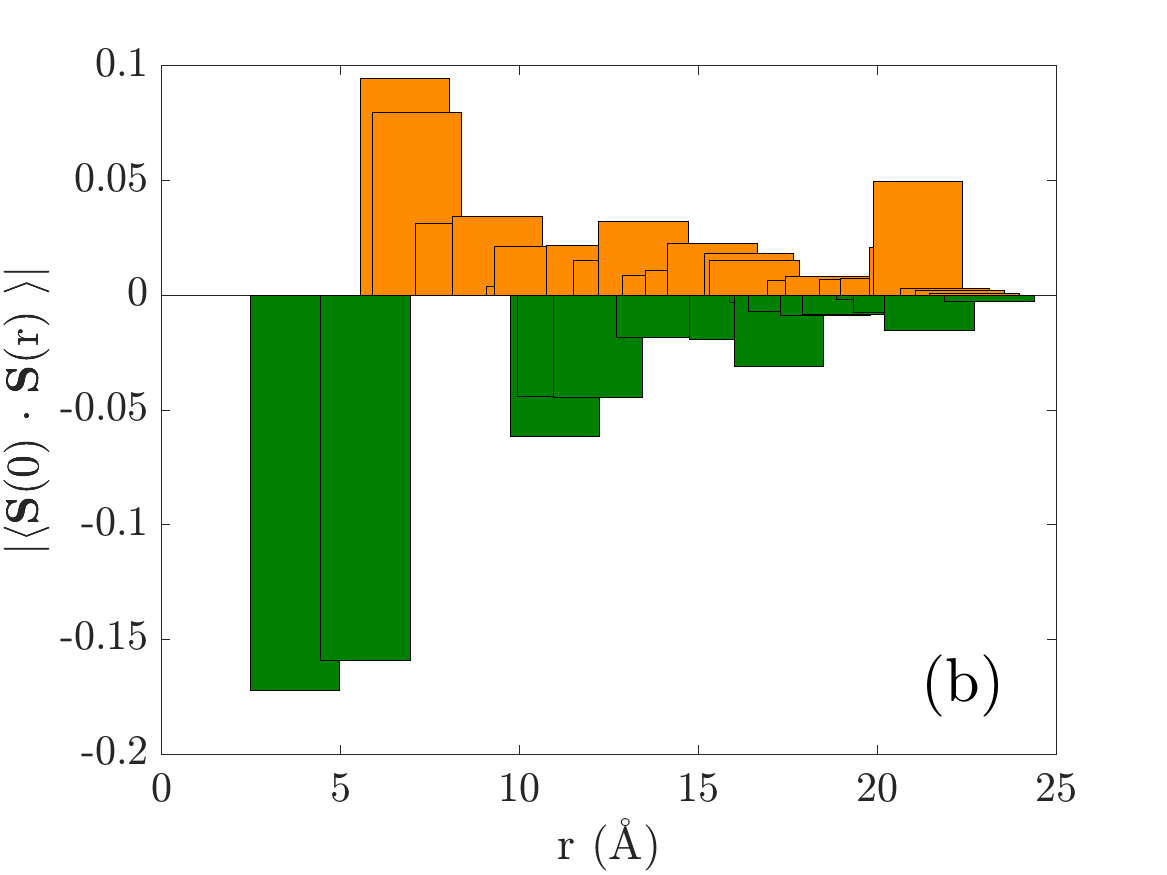}
	\caption{Correlation functions for MC simulations. (a) $J_1D$ model, (b) $J_1J_2D$ model. Positive scalars are coloured orange and negative scalars are coloured green.}
    \label{fig:MC:sisj}
\end{figure}

Figure \ref{fig:MC:C:Sq}(b) compares $S_{\rm mag}(\textbf{Q})$ and the resultant S(\textbf{Q}) for the $J_1D$ and  $J_1J_2D$ models. Both models provide features that are consistent with the data. The low $\textbf{Q}$ region is well reproduced by the $J_1D$ model, while this is not captured by the $J_1J_2D$ model. In contrast, the diffuse peaks at higher \textbf{Q} are reproduced by the $J_1J_2D$ model. These peaks do not appear in the $J_1D$ model. 

Figure \ref{fig:MC:sisj} presents the radial dependence of the spin-spin correlations for (a) the $J_1D$ model and (b) the $J_1J_2D$ model. Interestingly, the $J_1D$ model provides FM NN correlations, while the correlations across the loop are negative, and thus AFM. Correlations are not significant beyond the unit cell distance. The $J_1J_2D$ model has AFM NN correlations, with FM correlations across the ten-ion loop. In the $J_1J_2D$ model, the correlations remain significant for distances up to 25 \AA.

\section{Discussion}\label{sec:discussion}

We have studied the magnetically short ranged ordered state in YbGG, 0.1 $<$ T $<$ 0.6 K. We have revisited the crystal field excitations using inelastic neutron scattering and show that for  $T \leq 5$~K the Yb$^{3+}$ ions can be considered with an effective S = $1/2$ ion as spin-spin interactions dominate. In this description we obtain a negative Curie-Weiss temperature of $-0.2(1)$~K indicative of AFM interactions. Previous susceptibility measurements by Filippi et al. \cite{Filippi_1980} in the low temperature regime yielded a positive Curie-Weiss temperature of $+0.045(5)$~K indicative of FM interactions.Although these present inconsistent results, all susceptibility measurements agree that the spin-spin interactions are in the mK range and several orders of magnitude smaller than the crystal field energies.  

The magnetic scattering profiles, are determined in different manners from two different experiments: (1) via neutron polarisation analysis  $S_{\rm mag}(\textbf{Q})$ and (2) via a high temperature paramagnetic subtraction,$S_{\rm magff}(\textbf{Q})$. These two techniques provide different scattering profiles with PA providing an absolute magnetic scattering profile and the high temperature subtraction oversubtracting the formfactor contribution of the paramagnetic scattering. The datasets vary in \textbf{Q} and energy and \textbf{Q} and energy resolutions. We compare the energy resolved and energy integrated datasets directly since the static components of these datasets dominate. $S_{\rm mag}(\textbf{Q})$  and $S_{\rm magff}(\textbf{Q})$  presents short range correlated scattering with correlation lengths varying from near neighbour correlations to 20~\AA thereby providing confidence that we are probing the magnetically short ranged ordered regime. 

We have performed RMC simulations to extract the spin configurations from each data set considering the difference in $S_{\rm mag}(\textbf{Q})$ and $S_{\rm magff}(\textbf{Q})$. The RMC simulations, for three distinct datasets, provide spin configurations that are rather similar to each other but very distinct from the spin distribution found in the isostructural compound GGG. 

We compare the distribution of the azimuthal angle of the spins within a ten-ion loop and find that, for YbGG, each spin is peaked along the tangent of the loop, along the local z-direction, and with FM near neighbor correlations. In contrast, the spin structure for GGG provides a distribution perpendicular to the ten-ion loop. It is unclear what the origin of the significant anisotropy in YbGG along the local $z$-direction might be. Pearson et al.  \cite{Pearson_1967} calculated the diagonal elements of the crystal field $g$-factors and found these to be $g = (2.84, 3.59, -3.72)$, thus showing a slightly larger contribution along the local $z$-direction, but not significant to provide strong anisotropy. 

Extracting, the spin structure across the ten-ion loop, provides a director state with an easy axis along the local z-direction, comparable to the easy axis of the director state in GGG. The resultant director state of the ten-ion loop is also, similar to GGG, strongly anisotropic but unlike GGG is not long range ordered. 

We have studied a $J_1D$ and $J_1J_2D$ model using MC simulations for specific heat data and determined a range for the exchange interactions, $J_1$, $J_2$, and $D$. Both models providing convincing reproductions for the heat capacity data reproducing the short ranged ordered feature for 0.06 $<$ T $<$ 0.6 K and the long range lambda transition around 0.05 K. In GGG the relative $J_1/D$ value is $J_1/D = 0.107\; {\rm K} / 0.0457\; {\rm K} = 2.34$  while our MC simulations for YbGG yield $2.86\; (0.6\; {\rm K}/0.21\; {\rm K}) < J_1/D < 3.88\; (0.72\; {\rm K}/0.18\; {\rm K})$. 

The exchange interactions determined by MC are used to recalculate $S_{\rm mag}(\textbf{Q})$ using the relevant Hamiltonian for the $J_1D$ and $J_1J_2D$ models. The resultant scattering patterns are comparable, in part, to the experimental data. However our data and models provide no unique interpretation of the complete dataset. The $J_1D$ model, with ferromagnetic near neighbour spin-spin correlations, captures the low-$\textbf{Q}$ neutron scattering data while the $J_1J_2D$, with antiferromagnetic near neighbour spin-spin correlation, closely captures the data at higher $\textbf{Q}$. A more complex Hamiltonian is required to fully describe the magnetic state of YbGG and will be the focus of further studies. 

 \section{Conclusion}\label{sec:conclusion}
 
In conclusion, we have probed the enigmatic magnetic state of YbGG and have been able to deduce the magnetic correlations using a combination of RMC and MC to describe heat capacity and neutron scattering results. We derive the magnitude of the near neighbour exchange interactions $0.6$~K~$ < J_1 < 0.7$~K, $J_2 = 0.12$~K and the magnitude of the dipolar exchange interaction, $D$, in the range $0.18$~K~$ < D < 0.21$~K. 
Magnetic correlations develop below 0.6 K, in line with a broad feature in the specific heat data. Through RMC simulations we find a spin structure consistent with a director state, similar to that found in GGG for T $<$ 1 K with an associated broad feature in the specific heat. However, in YbGG, the director correlations are short ranged. The broad dataset cannot be fully described within the current, rather basic, model but provides an avenue for further studies. We welcome further elaborate insight.


\begin{acknowledgments}
	LS, RE, and IMBB were funded by Nordforsk through the NNSP project. This project was further supported by the Danish Agency for Science and Innovation though DANSCATT and Interreg. The work at the University of Warwick was funded by EPSRC, UK through Grant EP/T005963/1. This research used resources of the Spallation Neutron Source,a DOE Office of Science User Facility operated by Oak Ridge National Laboratory. We thank ILL and SNS for providing the facilities to perform the neutron scattering experiments. The authors would like to thank Prof. emeritus J. Jensen for valuable discussions. 
	\footnote{Note to the editor (not to be printed in the journal):
This manuscript has been authored by UT-Battelle, 
LLC under Contract No. DE-AC05-00OR22725 with the U.S. Department of Energy. 
The United States Government retains and the publisher, by accepting the 
article for publication, acknowledges that the United States Government 
retains a non-exclusive, paid-up, irrevocable, world-wide license to 
publish or reproduce the published form of this manuscript, 
or allow others to do so, for United States Government purposes. 
The Department of Energy will provide public access to these results 
of federally sponsored research in accordance with the DOE Public Access Plan 
(http://energy.gov/downloads/doe-public-access-plan).}

\end{acknowledgments}
\bibliography{YbGG_April2021}
\clearpage
\newpage
\appendix

\section{Crystal Field}\label{appendix_CF}
 The inelastic neutron scattering results from IN4 confirmed that YbGG has very strong crystal field levels and verified the excitation energies of the crystal field levels determined experimentally by Buchanan et al. \cite{Buchanan_1967} and theoretically by Pearson et al. \cite{Pearson_1967}.  In the calculations a crystal field Hamiltonian, 
\begin{align}
\mathcal{H}_{\text{CF}} &= \sum_i \sum_{lm} A_{lm} \langle r^l \rangle \alpha_l \left(\frac{2l+1}{4\pi}\right)^{1/2} \tilde{O}_l^m(J) \label{eq:HCF_A}\\ 
&= \sum_i \sum_{lm} B_l^m O_l^m(J),  \label{eq:HCF_B}
\end{align}
was optimised. Here, $\tilde{O}_l^m(J)$ are the Racah operators which transform like spherical harmonics, while $O_l^m(J)$ are the Stevens operators which transform like tesseral harmonics. $\alpha_l$ is the Stevens factor which depend on the form of the electronic charge cloud of the single ion, $A_{lm}$ is the effective charge distribution of the surrounding ions projected into the $Y_l^m$-basis, and $B_l^m$ are the Stevens parameters. Since both $\alpha_l$ and $\langle r^l \rangle$ are well defined from the system, there is direct correspondence between the $A_{lm}$ parameters and the $B_l^m$ parameters. 

Yb$^{3+}$ is a rare earth ion with 4f electrons as the outer shell. Consequently, $l\leq$7, but in order to obey time reversal symmetry, only even $l$ and $m$ are allowed, and the crystal symmetry excludes negative $m$. Consequently, there are 9 Stevens parameters with $l=2,4,6$ and $m\leq$l. Pearson et al. \cite{Pearson_1967} calculated the Stevens parameters using a point charge model approximation and later fitted the obtained parameters to experimental data of near-infrared spectroscopy and susceptibility measurements \cite{Buchanan_1967}. Table~\ref{tab:Stevens} shows the resulting Stevens parameters, which have been calculated based on the $A_{lm}$ parameters presented by Pearson et al.

\begin{figure}[!h]
	\centering
    \includegraphics[width = \linewidth]{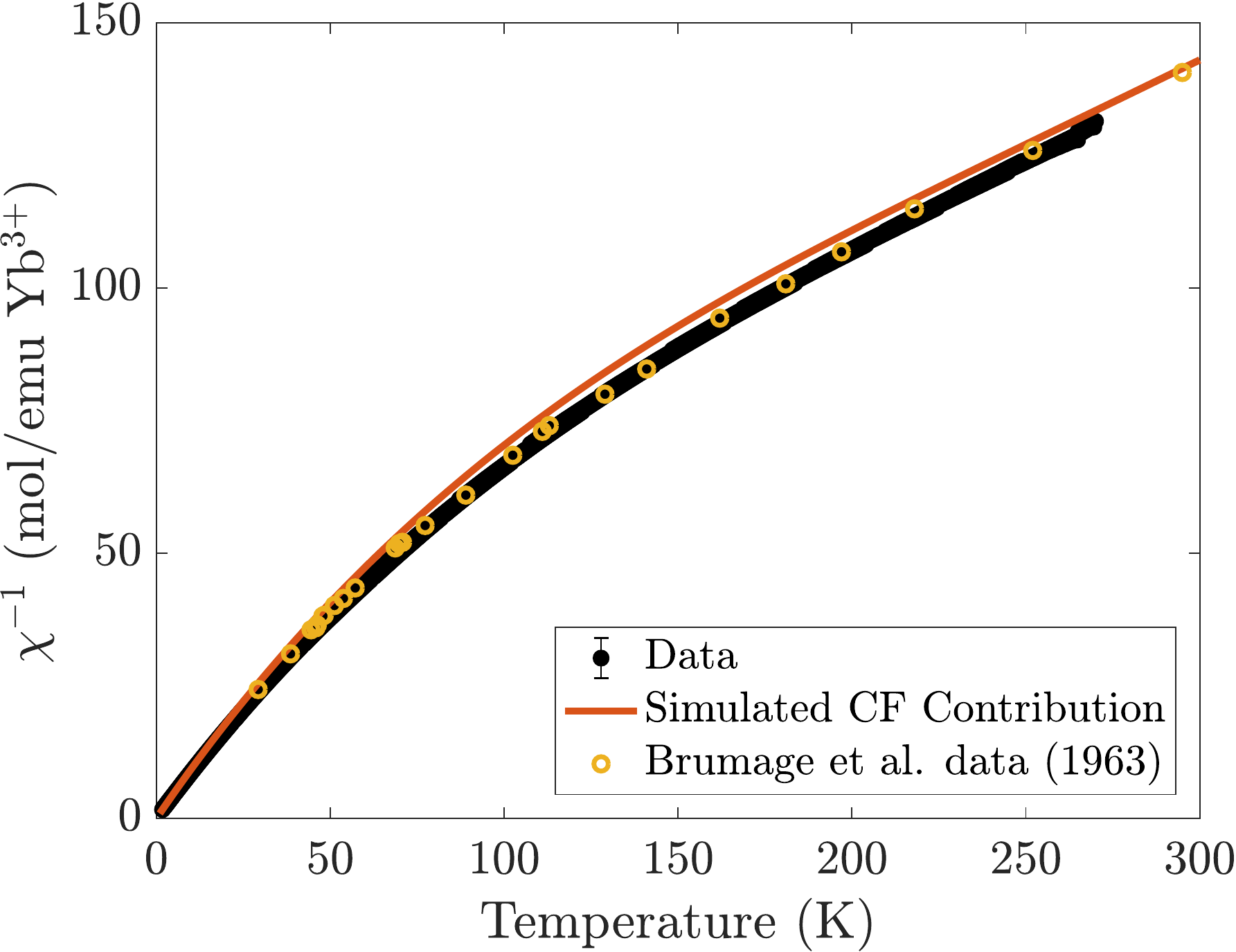}
	\caption{}\label{fig:CFsim_brumage}
\includegraphics[width = 0.5\linewidth]{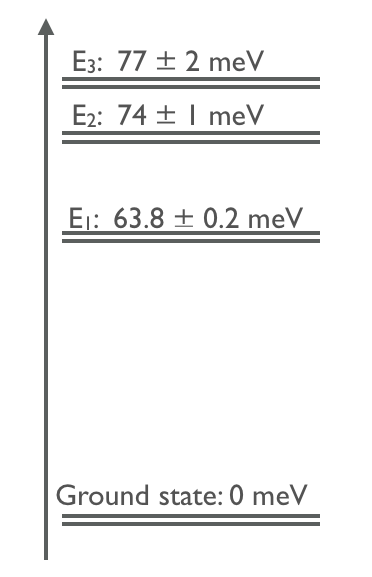}
\caption{(a) Susceptibility data, simulated  crystal field contribution along with previous susceptibility  measurements by Brumage et al.\cite{Brumage_1963} (b) Energy diagram of crystal field levels obtained from inelastic neutron scattering measurements on IN4.}
\label{fig:chi_Brumage}
\end{figure}

 The susceptibility has been simulated using the McPhase program with the Stevens parameters listed in Table~\ref{tab:Stevens}, using the values determined by Buchanan and Pearson \cite{Buchanan_1967, Pearson_1967}, without including any spin-spin interactions, such as exchange or dipolar interactions. Consequently, the simulated susceptibility, which is presented in Fig.~\ref{fig:chi_Brumage}(a) only contains the crystal field contribution to the susceptibility. The experimental data is well reproduced. It is thus possible to describe the susceptibility using only the crystal field considerations in the high temperature regime where the crystal field splitting is several orders of magnitude larger than the spin-spin interactions found from $\theta_{\rm CW}$. 


\begin{table}[!htp]	
\centering
	\begin{tabular}{|l|r|}
	\hline
	& Stevens parameters \cite{Pearson_1967} \\ \hline
	B$_{20}$ & -0.267 meV  \\
	B$_{22}$ & 1.097 meV  \\ \hline
	B$_{40}$ & 0.0368 meV  \\
	B$_{42}$ & -0.0459 meV  \\
	B$_{44}$ & -0.1291 meV  \\ \hline
	B$_{60}$ & 0.000870 meV  \\
	B$_{62}$ & -0.008205 meV  \\
	B$_{64}$ & 0.01460 meV  \\
	B$_{66}$ & -0.004138 meV \\ \hline 
	\end{tabular}
\caption{Stevens parameters obtained from Ref.~\onlinecite{Pearson_1967, Buchanan_1967}}
\label{tab:Stevens}
\end{table}

\begin{figure}[!h]
         \includegraphics[width = \linewidth]{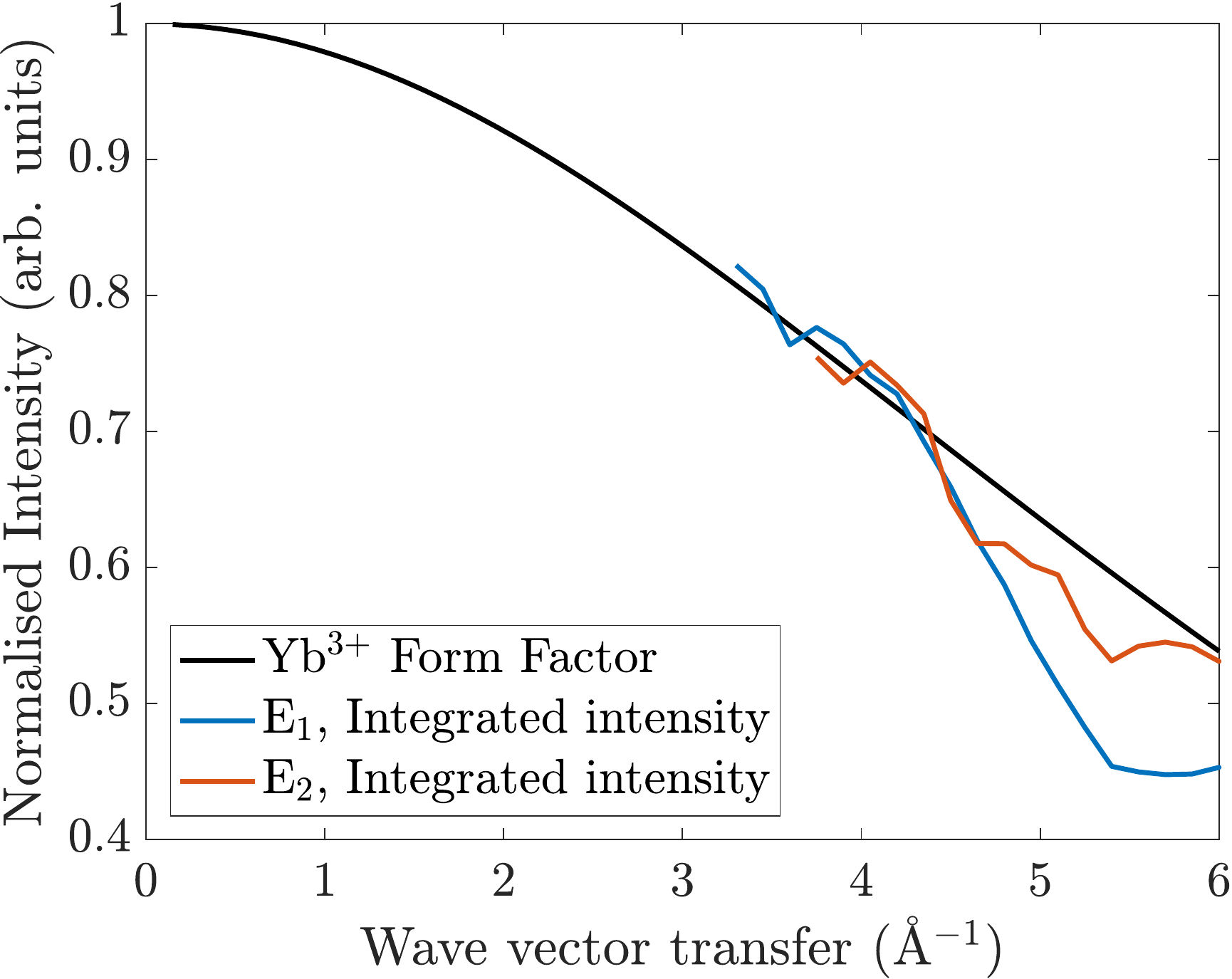}
    \caption{The integrated intensities of $E_1$ and $E_2$ from the IN4 measurements show good qualitative agreement with the calculated form factor of Yb$^{3+}$.}
\label{fig:IN4_Fits}
\end{figure}

\FloatBarrier
\section{Crystal}
A single crystal of YbGG has been grown using the floating zone method in Ar + O$_2$ gas mixture at a growth rate of 10 mm/h\, \cite{Kimura_2009}. The results achieved thus far indicate that the crystals quality and size are suitable for studies of magnetic frustration studies using neutron diffraction. Synthesizing crystals with an adequate volume for neutron scattering is complex due to the weak scattering cross sections and thus the requirement for large (cm$^{3}$) single crystals. As such, the growth of a large single crystal is a success. X-ray Laue diffraction after growth determined sample crystallinity and orientation. Future work will include a detailed analysis of the effects of stoichiometry, vacancies and site mixing on the magnetic behaviour of YbGG garnets. 
\begin{figure}[!h]
	\centering
    \includegraphics[width = 0.7\linewidth]{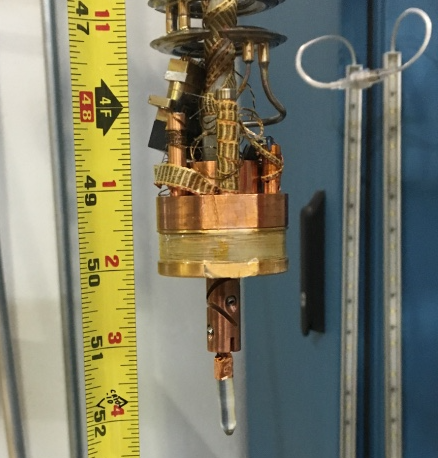}
	\caption{Crystal used for neutron scattering experiments and susceptibility measurements}
    \label{fig:YbGGPhoto}
\end{figure}

\FloatBarrier
\section{Elastic neutron scattering data and linecuts}\label{app:SqData}
This section contains elastic 2D data along with Gaussian fits of linecuts through the elastic neutron scattering data to quantify the observed diffuse features. 

\begin{figure}
    \centering
     \includegraphics[width = \linewidth]{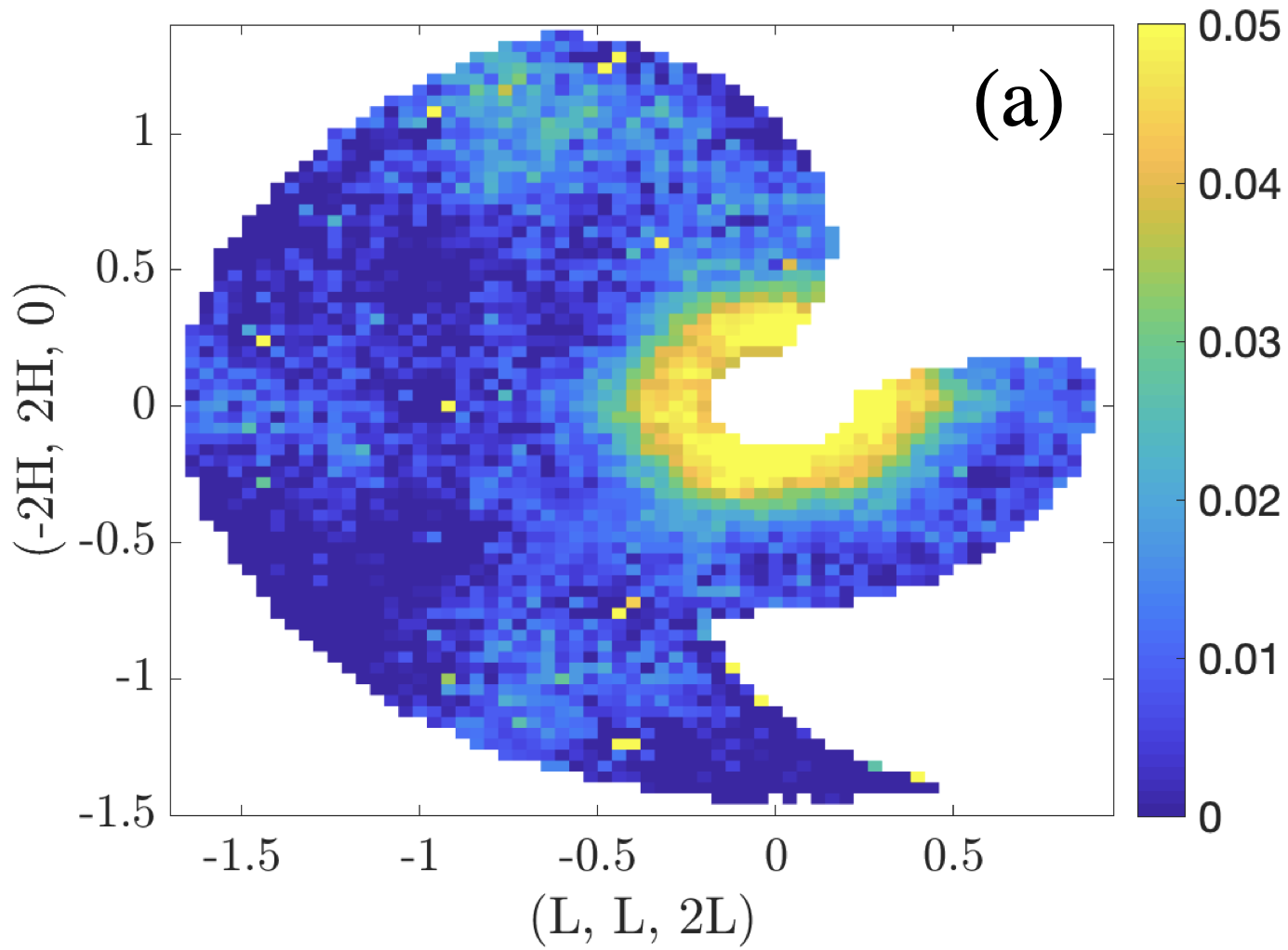}
     \includegraphics[width = \linewidth]{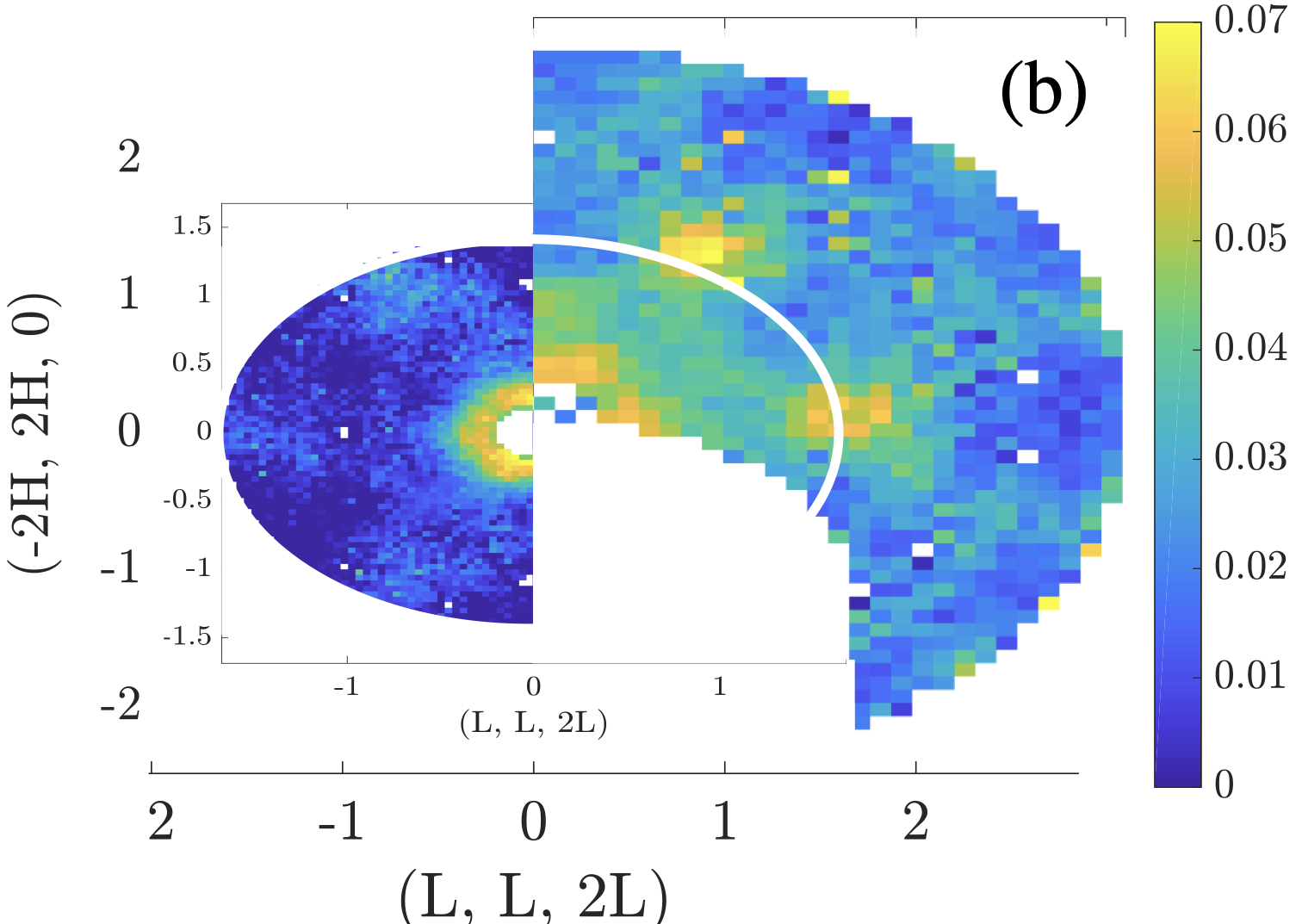}

    \caption{(a)Magnetic contribution to $S(Q,E=0)$, measured at CNCS with $E_{\rm i} = 3.32$~meV.(b) The relative reciprocal space accessed for the CNCS, $E_{\rm i} = 3.32$~meV data set and the D7 data set. }
    \label{fig:3p32_2Ddata}
\end{figure}


\begin{table}[htb]	
\centering
	\begin{tabular}{|l|ll|}
	\hline
	$E_{\rm i} = 1.55$~meV &&\\
	\hline
	Gaussian Peak Position & 0.30 &$\pm$ 0.03 \AA$^{-1}$ \\
    Distance (from peak pos) & 20 &$\pm$ 2 \AA \\
    FWHM & 0.52 &$\pm$ 0.09 \AA$^{-1}$ \\
    Correlation Length (from FWHM) & 12 &$\pm$ 2 \AA
    \\ && \\ \hline 
	$E_{\rm i} = 3.32$~meV && 
	\\ \hline
	Gaussian Peak Position & 1.86 &$\pm$ 0.09 \AA$^{-1}$\\
    Distance (from peak pos) & 3.4 &$\pm$ 0.2 \AA \\
    FWHM & 0.55 &$\pm$ 0.25 \AA$^{-1}$ \\
    Correlation Length (from FWHM) & 11.5 &$\pm$ 5.2 \AA 
	\\ && \\ \hline
	$E_{\rm i} = 8.11$~meV && 
	\\ \hline
	Gaussian Peak Position & 1.95 &$\pm$ 0.08 \AA$^{-1}$ \\
    Distance (from peak pos) & 3.2 &$\pm$ 0.1 \AA \\
    FWHM & 1.6 &$\pm$ 0.2 \AA$^{-1}$ \\
    Correlation Length (from FWHM) & 4.0 &$\pm$ 0.6 \AA
    \\ && \\ \hline
	\end{tabular}
\caption{Fit parameters of the 3 Gaussian fits in Fig.~\ref{fig:LineCutsGaussianFits}.}
\label{tab:GaussFit}
\end{table}
Figure \ref{fig:3p32_2Ddata} show the 2D $S(\textbf{Q}, E=0)$ data obtained from the CNCS measurements with $E_{\rm i} = 3.32$~meV. Fig.~\ref{fig:3p32_2Ddata} contains only the magnetic contribution derived by subtracting a 13~K data set from a 0.05~K dataset. The signal to noise-ratio in the data is lower than the two other elastic neutron scattering datasets presented in the main text. 

\begin{figure}[!h]
    \includegraphics[width = \linewidth]{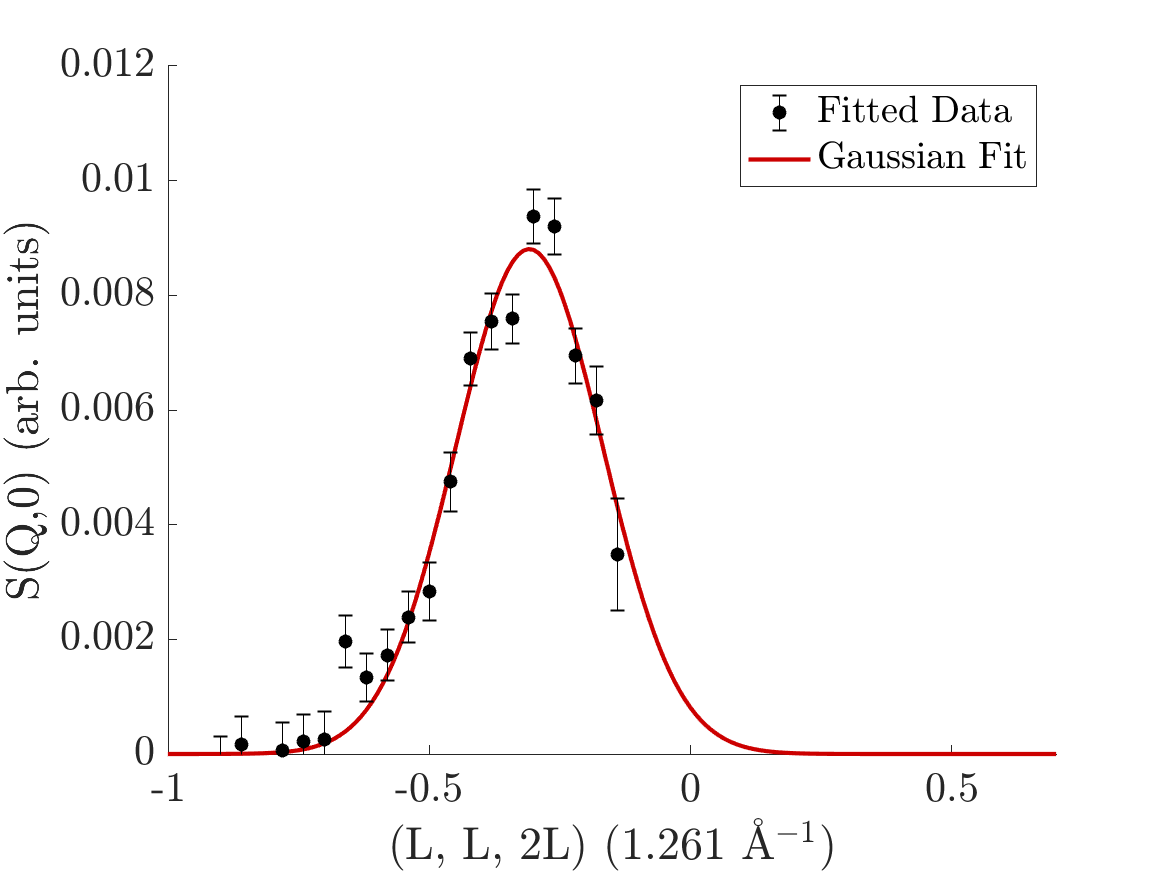}
    \includegraphics[width = \linewidth]{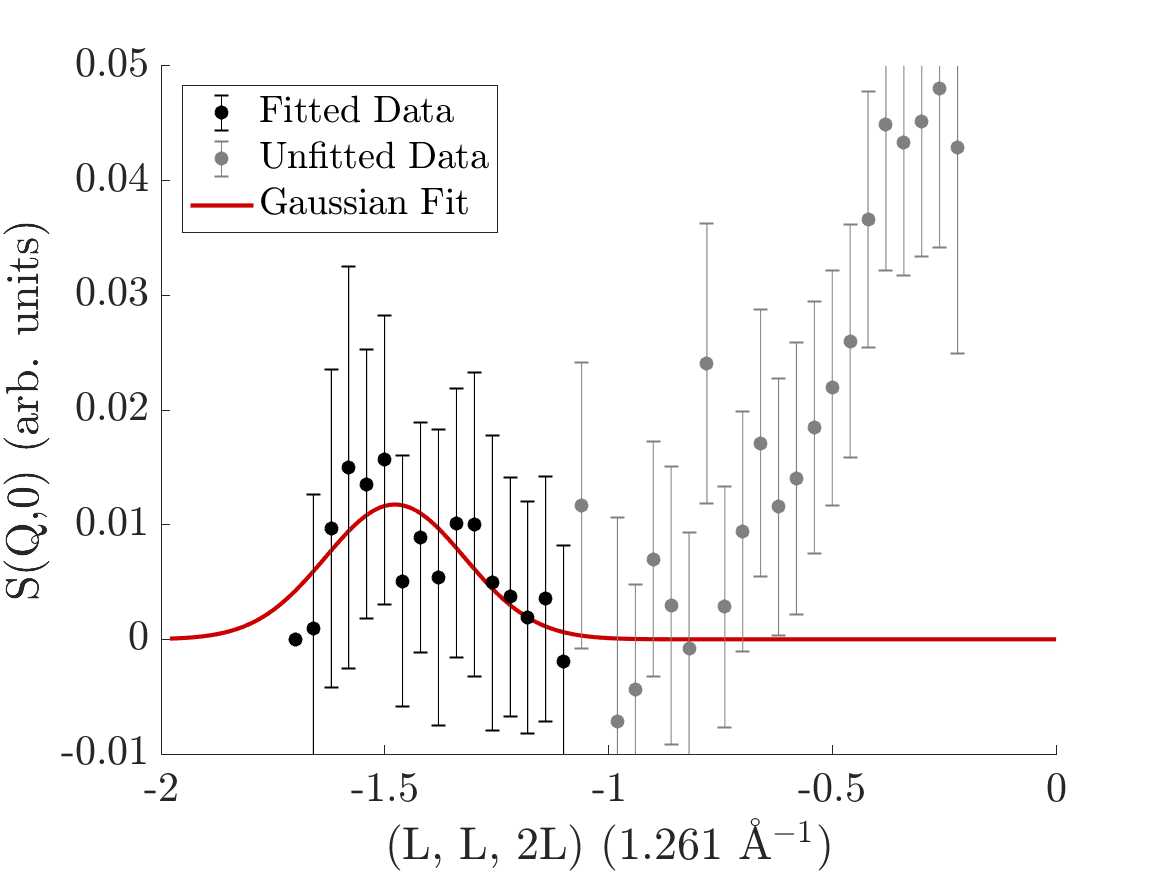}
    \includegraphics[width = \linewidth]{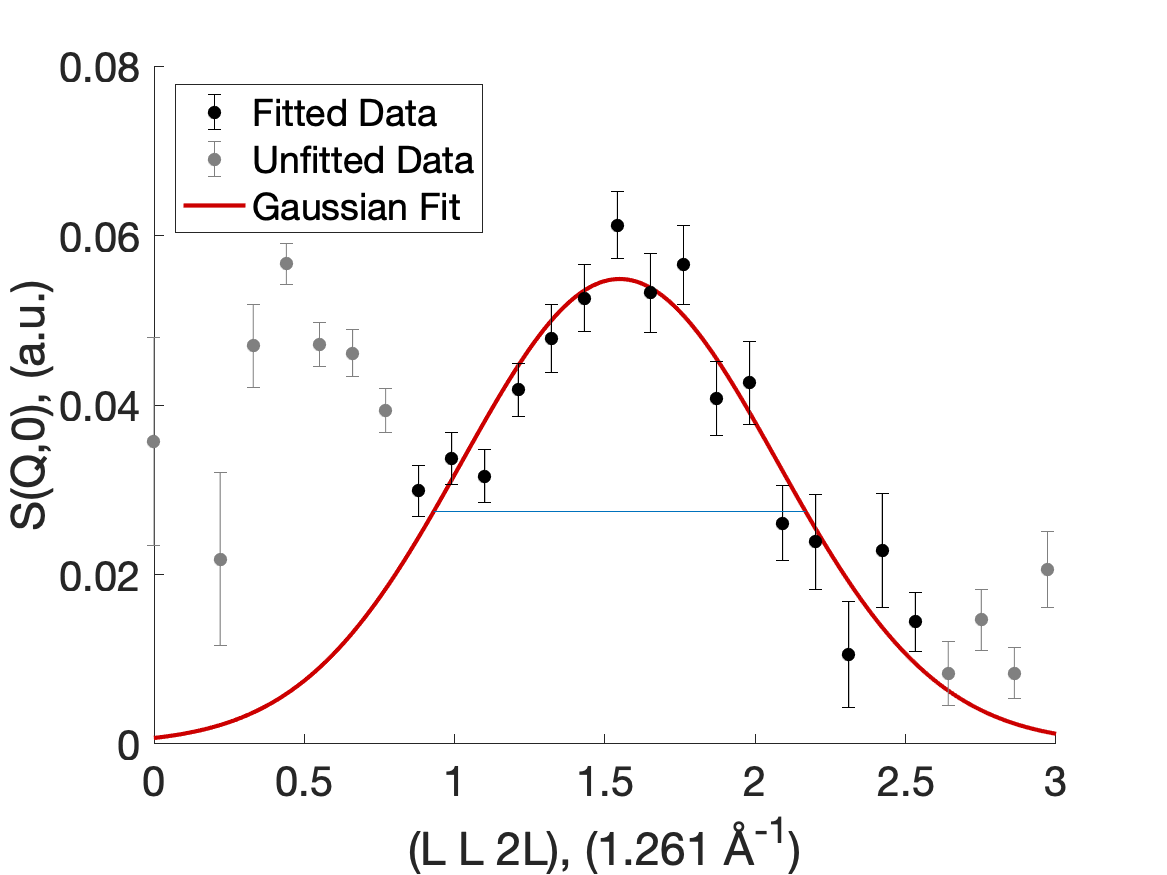}
    \caption{Line cuts from the two-dimensional neutron scattering data together with Gaussian fits. (a) CNCS data, $E_{\rm i} = 1.55$~meV. (b) CNCS data, $E_{\rm i} = 3.32$~meV. (c): D7 data, $E_{\rm i} = 8.11$~meV. Fit parameters are presented in table \ref{tab:GaussFit}.}
	\label{fig:LineCutsGaussianFits}
\end{figure}

This is supported by the data in Fig.~\ref{fig:LineCutsGaussianFits}, which
shows various line cuts from the two-dimensional neutron scattering data together with Gaussian fits.

Fig.~\ref{fig:LineCutsGaussianFits}(a) shows a Gaussian fit to a linecut through the CNSC data with $E_{\rm i}=1.55$~meV, where (-2H 2H 0) = (0 0 0). Low Q hexagon peaks are seen at $\vert Q \vert = 0.30 \pm 0.03 $~\AA$^{-1}$, corresponding to an equivalent magnetic lattice spacing of $d=20 \pm 2 $~\AA. The correlation length, obtained by the FWHM$ = 0.41 \pm 0.07$~\AA$^{-1}$, becomes $12\pm 2 $~\AA. 

Fig.~\ref{fig:LineCutsGaussianFits}(b) shows a Gaussian fit to a linecut through the CNCS data with $E_{\rm i}=3.32 $~meV, where (-2H 2H 0) = (0 0 0). The Gaussian Peak Position is $\vert Q \vert = -1.86 \pm 0.09$~\AA$^{-1}$, giving an equivalent lattice spacing of $d = 3.4 \pm 0.2$~\AA. The FWHM is $0.43 \pm 0.20$~\AA$^{-1}$, giving a correlation length of $12 \pm 5$~\AA.

Fig.~\ref{fig:LineCutsGaussianFits}(c) shows a Gaussian Fit to a linecut in the D7 data, where (-2H 2H 0) = (0.1 0.1 0). The Gaussian Peak Position is $3.14 \pm 0.13$~\AA$^{-1}$, giving an equivalent lattice spacing of $d = 2.0 \pm 0.1$~\AA. The FWHM is $1.89 \pm 0.29$~\AA$^{-1}$, giving a correlation length of $2.6 \pm 0.4$~\AA.

\FloatBarrier
\newpage 
\section{RMC refinements}\label{app:RMC}
\subsection{Method and additional data}
We follow the procedure of the \textit{Spinvert} refinement program\cite{Paddison_2013_RMC} and use a Monte Carlo technique to find classical Heisenberg spin configurations that can reproduce the experimentally observed scattering pattern. In theory, the spin-spin correlations $\langle \textbf{S}^{\perp}_i\cdot \textbf{S}^{\perp}_j \rangle$ are uniquely related to the magnetic scattering intensity. For clarity, we shall in this appendix use $\left( \frac{d\sigma}{d \Omega}\right)$ for the experimental signal and $S$ for the theoretically calculated signal from a single configuration. Assuming that we can describe the observed scattering with a static Heisenberg spin configuration, we are interested in the set of $M$ equations
\begin{equation}\label{Richard:ScatteringLawEquationSystem}
\begin{split}
    \left\{\left( \frac{d\sigma}{d \Omega}\right) (\textbf{Q}_k) =S\left(\{\textbf{S}_i\}_{i=1}^N,\textbf{Q}_k\right) \right\}_{k=1}^{M},\\
    S \equiv \frac{C[f(|\textbf{Q}|)]^2}{N}\sum_{i, j }\left\langle \textbf{S}^{\perp}_i\cdot \textbf{S}^{\perp}_j\right\rangle e^{i\textbf{Q}\cdot \textbf{r}_{ij}},
\end{split}
\end{equation}

which relates a spin configuration $\{\textbf{S}_i\}_{i=1}^N$ of $N$ spins to the scattering intensity. $k$ is a labeling index for all allowed $\{\textbf{Q}_k\}_{k=1}^M \subset \mathbb{R}^3 $ points. Ideally, the refinement method uses knowledge of the experimental left hand side of this system of equations to compute $\{\textbf{S}_i\}_i^N$. In particular we use single spin flips in simulated annealing to minimize the residual

\begin{equation}\label{Richard:Eq:residualVector}
    \chi^2 \equiv \sum_k \left(\left( \frac{d\sigma}{d \Omega}\right) (\textbf{Q}_k) - S\left(\{\textbf{S}_i\}_{i=1}^N,\textbf{Q}_k\right)   \right)^2.
\end{equation}

The experiment only gives information about $\textbf{Q}$-points in the $(-2H, 2H, 0)$, $(L,L,2L)$ plane and in the following we shall discuss what can be deduced about the underlying configurations.
We find that refining a solution only to the plane where the data was taken ends up over-fitting scattering intensities at unconstrained $\textbf{Q}$-points outside the plane, giving unphysical results. We made several attempts to compensate for this, such as adding mirrors of the plane in different directions allowed by the crystal symmetries to try and capture more of $\textbf{Q}$-space. However this was is not enough to resolve the issue. We conclude that with an under-constrained set of equations we will always over-fit in the simulated annealing and do not find physical solutions which are continuous and respect the crystal symmetries. 

Hence, we investigate possible ways of fully constraining the set of equations given the data. We need to postulate a scattering intensity for every $\textbf{Q}$-point, in order to avoid over-fitting. Since we do not have information about scattering intensities outside the measured plane, our attempt will be to extrapolate from the data the scattering intensities outside the plane to achieve a refinement result that agrees with the measured data and is continuous in the rest of $\textbf{Q}$-space. Naturally we can not assume to get a correct description of the spin configurations if we do not have access to the full diffraction pattern. Hence, we accept that the data presented is only an approximation of the true correlations. However, by testing several extrapolation techniques in addition to extracting RMC from various data sets with different incident energies and averaging across 400 RMC minimizations, we believe that the results are stable and that some variation in the assumed extrapolation will not affect the fundamental structure of the solution.

\begin{figure}[!h]
	    \includegraphics[width=\linewidth]{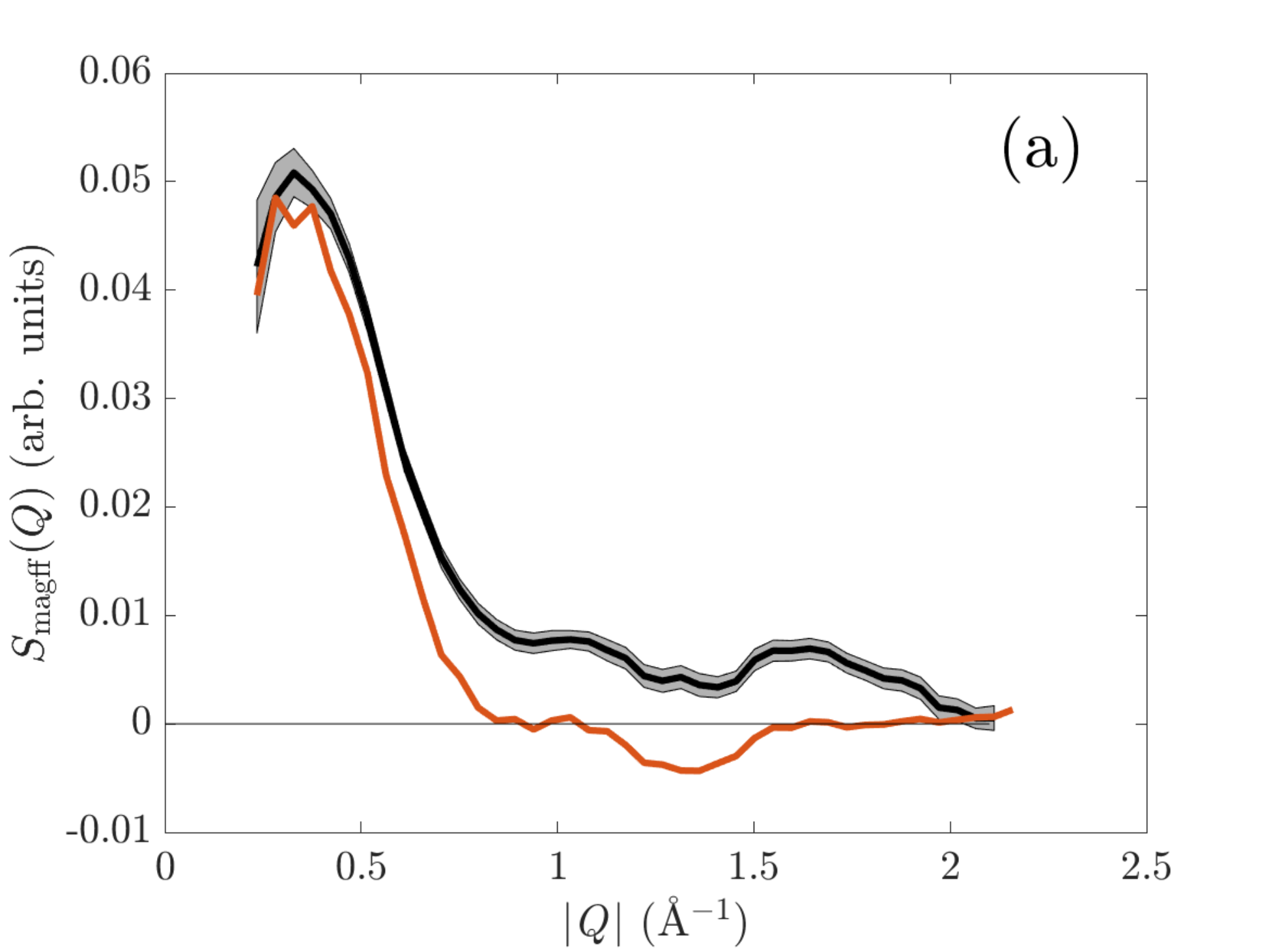}
	    \includegraphics[width=\linewidth]{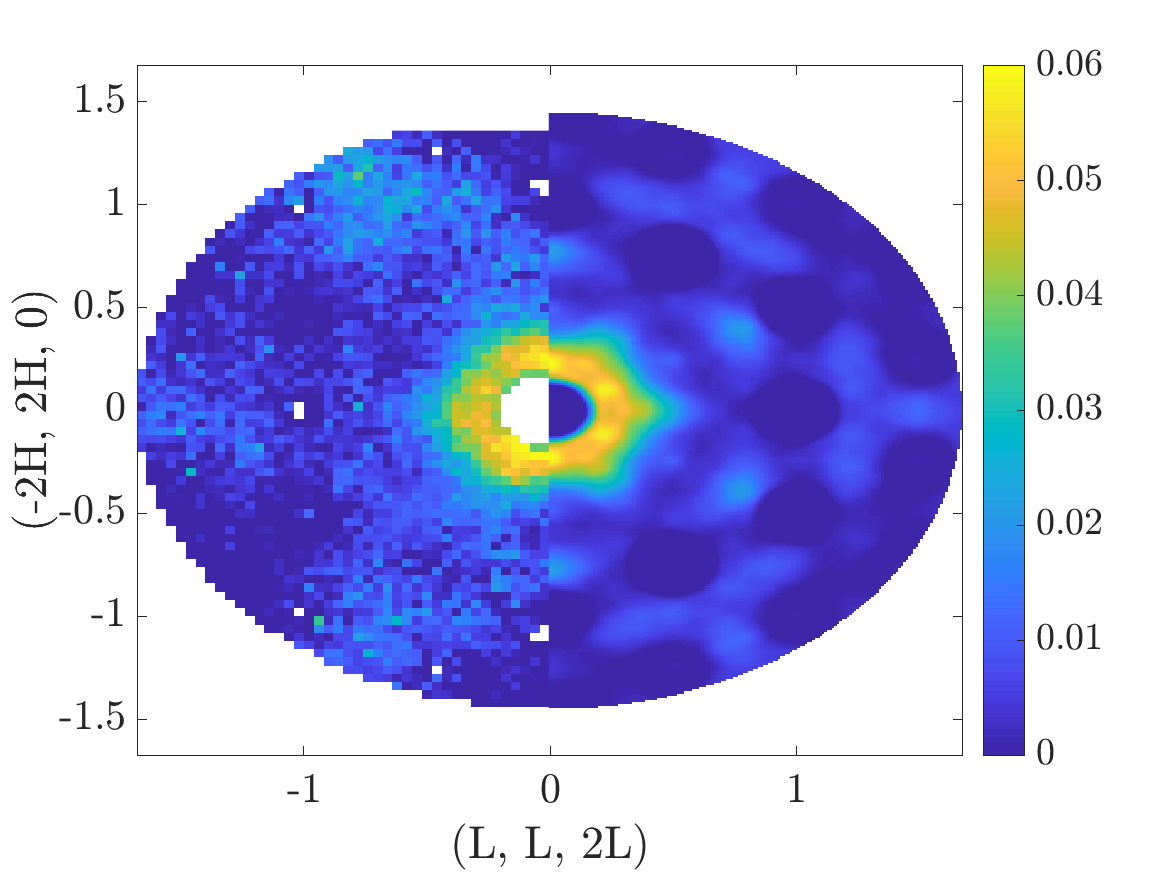}
	\caption{(a) The constructed powder average from the CNCS, $E_{\rm i} = 3.32$~meV data set $(\frac{d \sigma }{d \Omega})_\textup{magff}(Q)$ (black), together with the RMC fit $S_\textup{magff}(Q)$ (red). (b) The CNCS, $E_{\rm i} = 3.32$~meV data set (left) together with the RMC average $S_\textup{magff}(\textbf{Q})$ over 400 configurations (right). }
	\label{Richard:fig:RMCrefinementResults_total}
\end{figure}

To construct a three dimensional data set for the scattering intensity, we make the assumption that the scattering has the same directional average for a given $Q=|\textbf{Q}|$ in the experimentally measured plane as it has over all directions. The open source available Spinvert program\cite{Paddison_2013_RMC} is built for refining scattering data from powder samples by transforming Eq.(\ref{Richard:ScatteringLawEquationSystem}) into a powder average that depend only on $Q$, Eq.~(\ref{Richard:Eq:powderScattering}). We term the calculated powder average $S(Q)$ (as opposed to $S(\textbf{Q})$) for which we minimize the residual against the constructed powder average $\left(\frac{d\sigma}{d \Omega}\right)(Q)$, named the \textit{powder diffraction pattern} in the main text, and defined as

\begin{equation}
    \left(\frac{d \sigma}{d \Omega}\right)(Q) \equiv \frac{1}{\mathcal{M}(Q)}\sum_{||\textbf{Q}_k|-Q|<t} \left(\frac{d \sigma}{d \Omega}\right)(\textbf{Q}_k).
\end{equation}
where $\mathcal{M}(Q)$ is the number of $\textbf{Q}$-points in the experiment of magnitude $Q\pm t$. We choose the tolerance $t$ so that features can still be resolved and that good statistics are obtained. For the D7 data, we have the magnetic signal, denoted by the subscript ``$\textup{mag}$", from the experiment and we directly minimize the residual

\begin{equation}\label{Richard:Eq:residualMag}
    \chi_{\textup{mag}}^2 \equiv \sum_k \left(\left( \frac{d\sigma}{d \Omega}\right)_\textup{mag} (Q_k) - S_\textup{mag}\left(\{\textbf{S}_i\}_{i=1}^N,Q_k\right)   \right)^2.
\end{equation}

For the CNCS data, we obtain the magnetic signal as the subtraction of the $13\textup{ K}$ paramagnetic signal from the $0.05{\textup{ K}}$ signal. We use the subscript ``$\textup{magff}$" to indicate this and minimize the residual 

\begin{equation}\label{Richard:Eq:residualTS}
    \chi_{\textup{magff}}^2 \equiv \sum_k \left(\left( \frac{d\sigma}{d \Omega}\right)_\textup{magff} (Q_k) - S_\textup{magff}\left(\{\textbf{S}_i\}_{i=1}^N,Q_k\right)   \right)^2.
\end{equation}

$S_\textup{mag}(Q)$ and $S_\textup{magff}(Q)$ are given by
\begin{equation}\label{Richard:Eq:powderScattering}
\begin{split}
     &S_{\textup{magff}}(Q)=sC[\mu F(Q)]^2 \frac{1}{N}\times\\ &\sum_{i,j}\left[A_{ij}
     \frac{\sin Qr_{ij}}{Qr_{ij}}+B_{ij}\left(\frac{\sin Qr_{ij}}{(Qr_{ij})^3}-\frac{\cos Qr_{ij}}{(Qr_{ij})^2}\right)\right],\\
     &S_\textup{mag}(Q)=S_{\textup{magff}}(Q)+\frac{2sC}{3N}[\mu F(Q)]^2 
\end{split}
\end{equation}
where
\begin{equation}
\begin{split}
    A_{ij}=\textbf{S}_i\cdot\textbf{S}_j-(\textbf{S}_i\cdot\hat{\textbf{r}}_{ij})(\textbf{S}_j\cdot\hat{\textbf{r}}_{ij}),\\
    B_{ij}=3(\textbf{S}_i\cdot\hat{\textbf{r}}_{ij})(\textbf{S}_j\cdot\hat{\textbf{r}}_{ij})-\textbf{S}_i\cdot\textbf{S}_j.\\
\end{split}
\end{equation}
$F(Q)$ is the magnetic form factor of $\textup{Yb}^{3+}$, $\mu$ is the effective dipole moment of $\textup{Yb}^{3+}$, $C=0.07265 \textup{ barn}$ is a physical constant. $N$ is the number of particles in the refinement supercell and $s$ is an overall dimensionless scale factor which relates neutron counts to the differential cross section. Due to the complexity of determining this scale factor, we choose to probe the solution space for all values of $s$. The resulting refinement will depend in a non-trivial way on $s$ and from the subset of configurations that minimize the residual of Eq.~(\ref{Richard:Eq:residualMag}) or Eq.~(\ref{Richard:Eq:residualTS}) we determine the best fit from the residual of Eq.~(\ref{Richard:Eq:residualVector}), where the directional dependence is included. From this definition of the best fit, we take the average of 400 minimizations to obtain the RMC fits presented in the main text. Here, we also present the RMC fit to the secondary CNCS data set, Fig.~\ref{Richard:fig:RMCrefinementResults_total}.

\begin{figure}[!h]
	\centering
	    \includegraphics[width=0.45\linewidth]{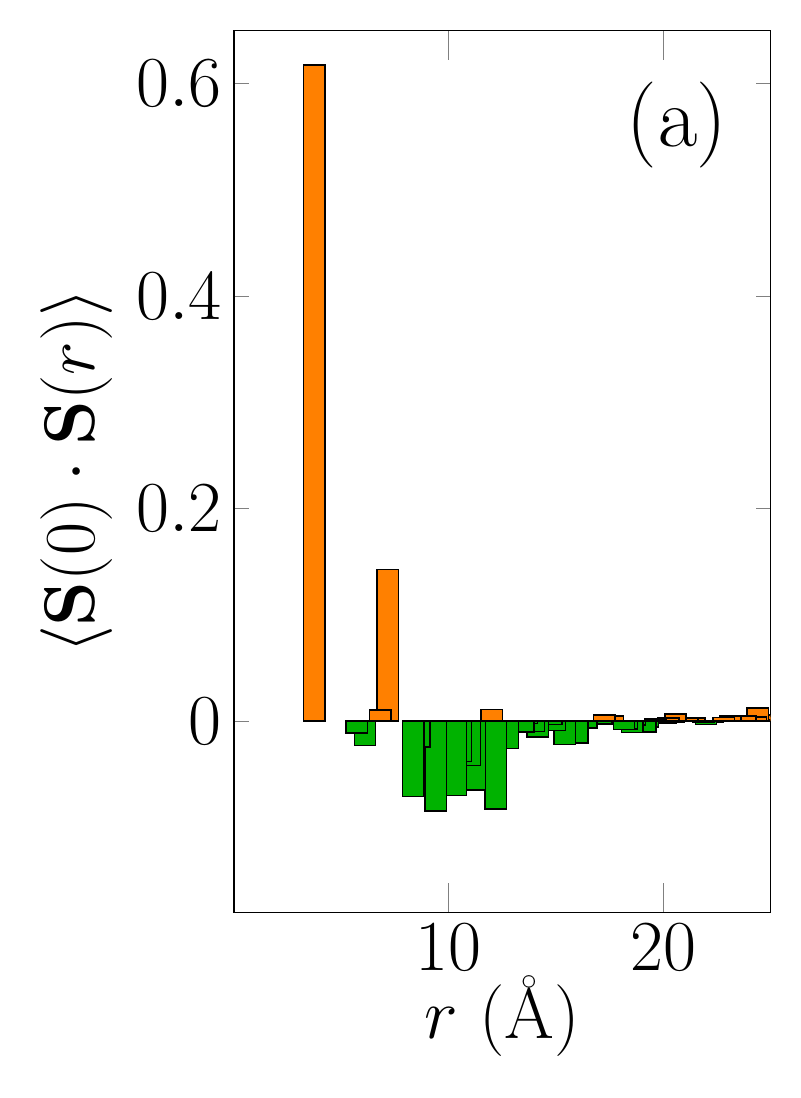}
	    \includegraphics[width=0.35\linewidth]{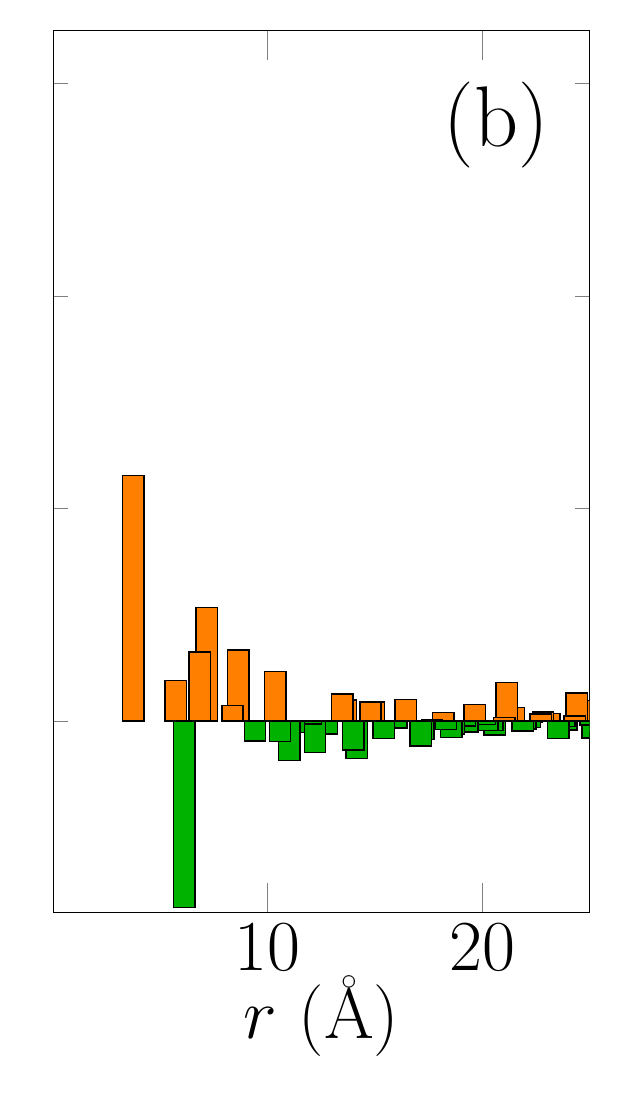}
	    \includegraphics[width=0.45\linewidth]{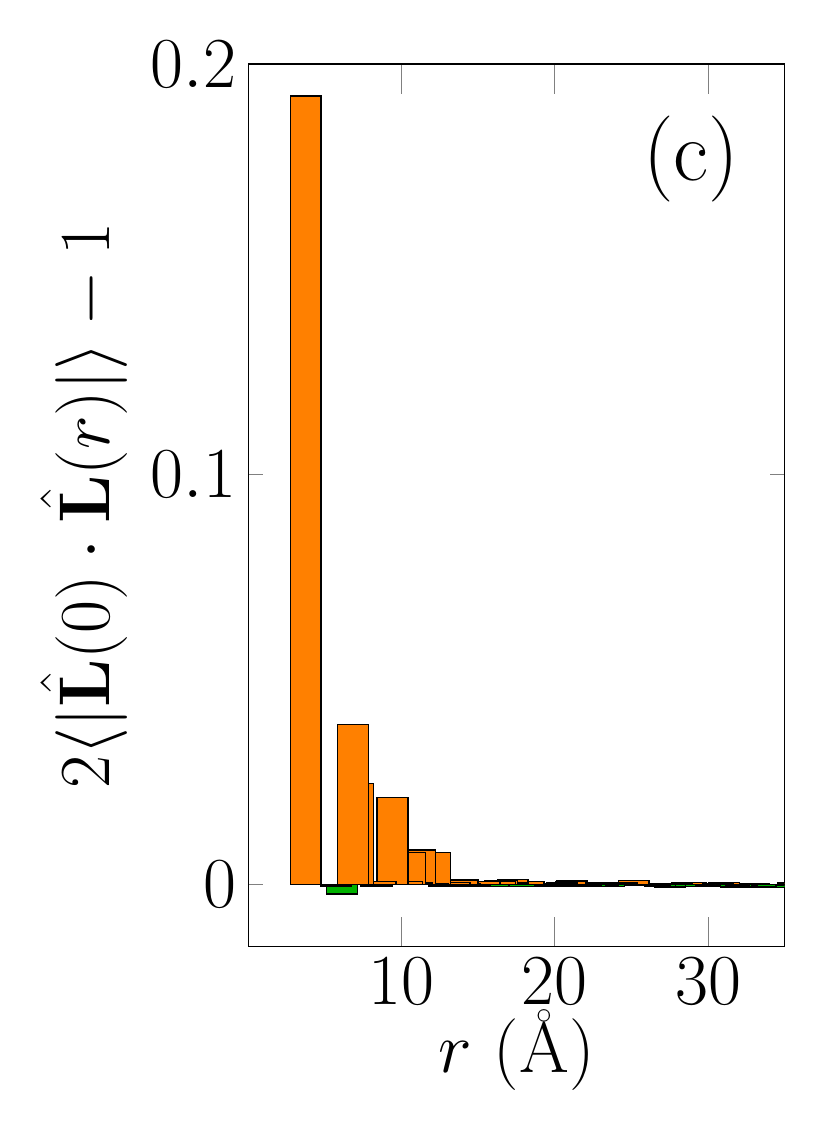}
	    \includegraphics[width=0.35\linewidth]{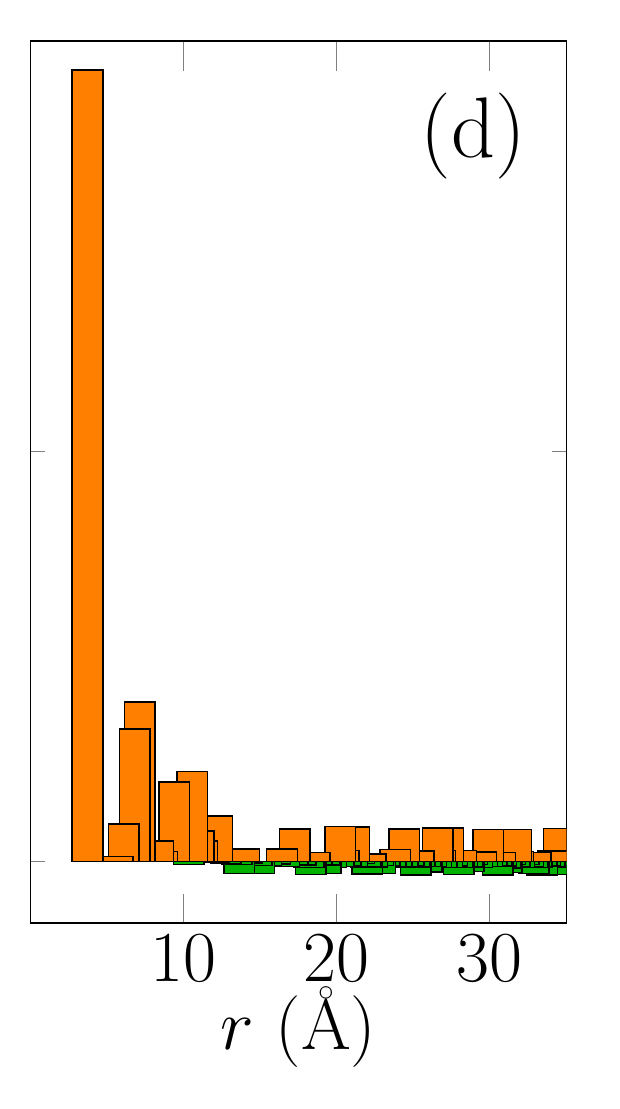}
	\caption{Spin-spin correlation function and director-director correlation function from the RMC refinements of the other data sets. (a) and (b) show the spin-spin correlations for the CNCS $E_{\rm i} =3.32 \textup{MeV}$ and D7 refinements respectively. (c) and (d) show director-director correlation function, $\langle|\hat{\textbf{L}}(0)\cdot\hat{\textbf{L}}(r)|\rangle-1$. ($g_L(r)$) for CNCS $E_{\rm i} =3.32$~meV and the D7 data respectively.
	}\label{Richard:fig:SiSjAndLiLj1p55}
\end{figure}

In Fig.~\ref{Richard:fig:SiSjAndLiLj1p55} we show the RMC spin-spin and director-director correlation functions for the D7 and CNCS $E_{\rm i} =3.32 \textup{ meV}$ data sets, which we left out in the main text. We see that the average product between nearest neighbour spins is positive, just as in the main text.

\begin{figure}[!h]
    \centering
	    \includegraphics[width=0.45\linewidth]{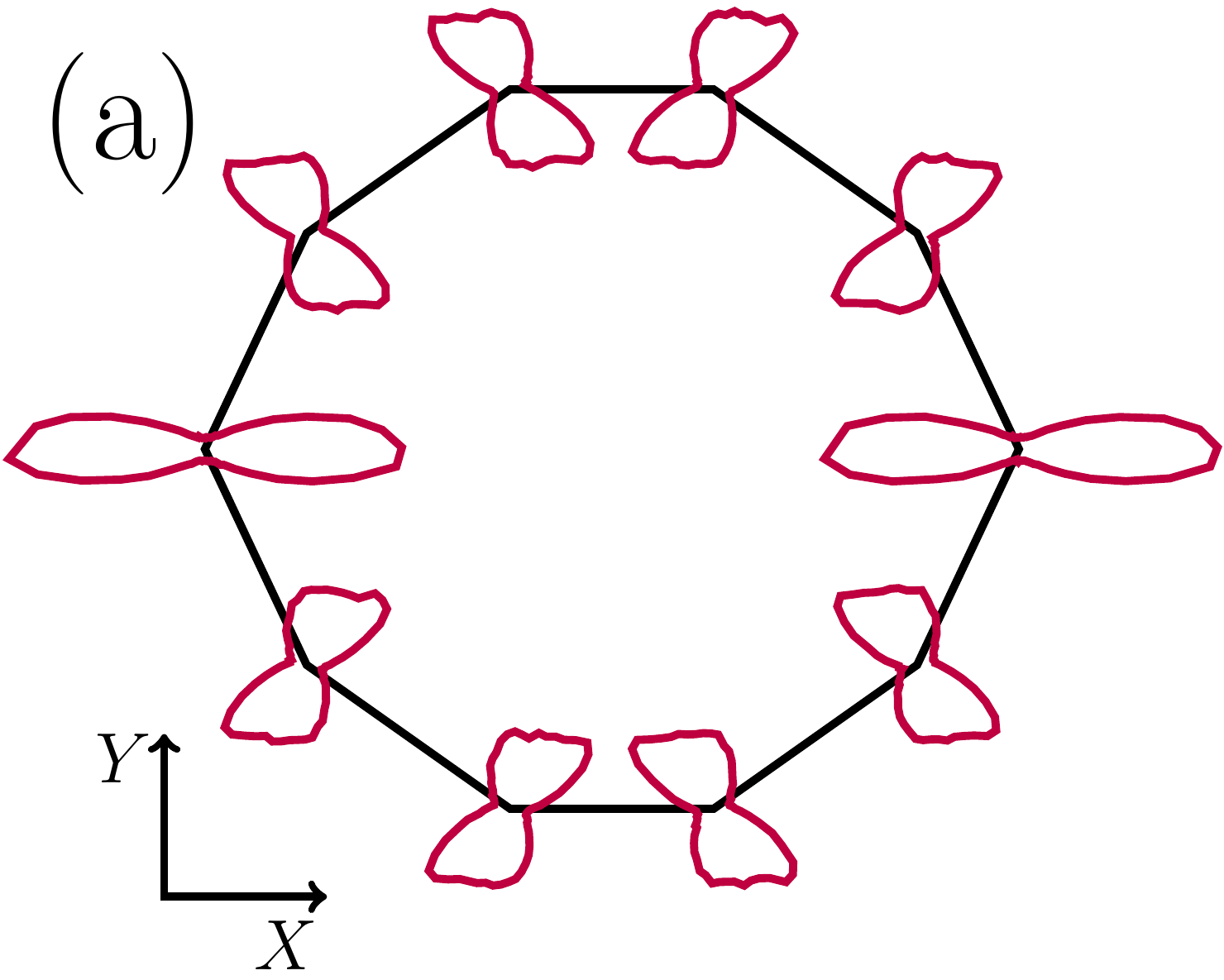}
	    \includegraphics[width=0.45\linewidth]{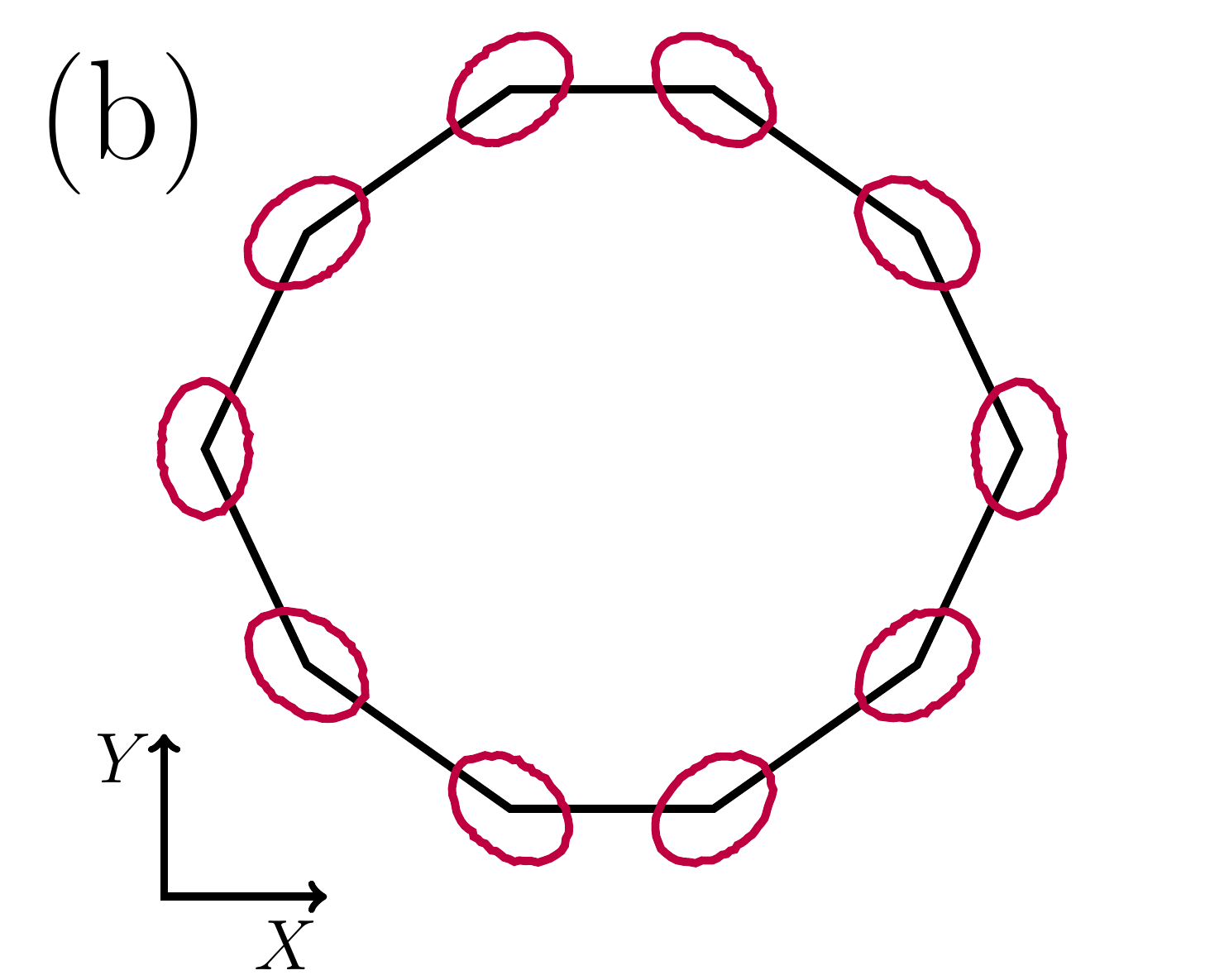}
	    \includegraphics[width=0.45\linewidth]{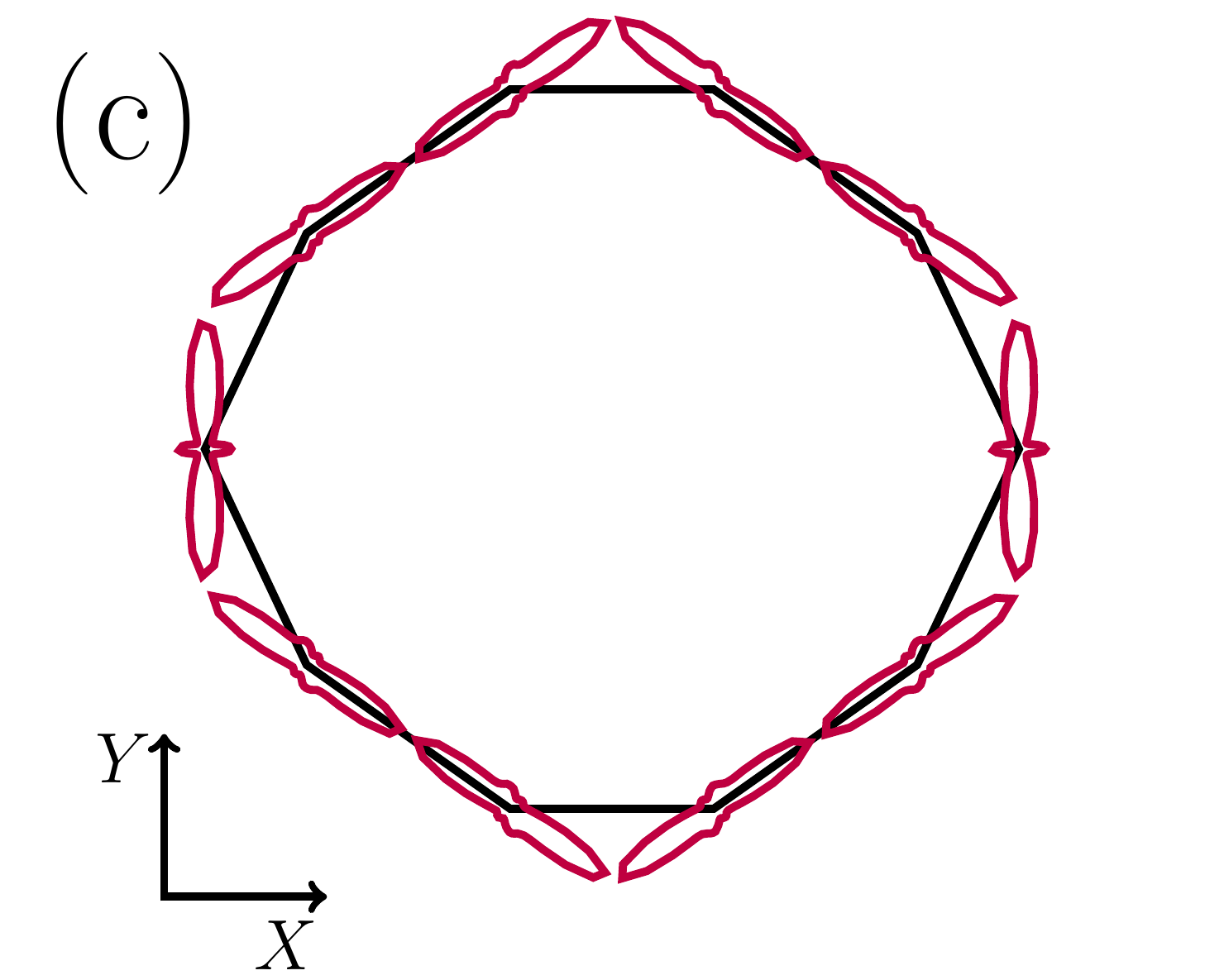}
	    \includegraphics[width=0.45\linewidth]{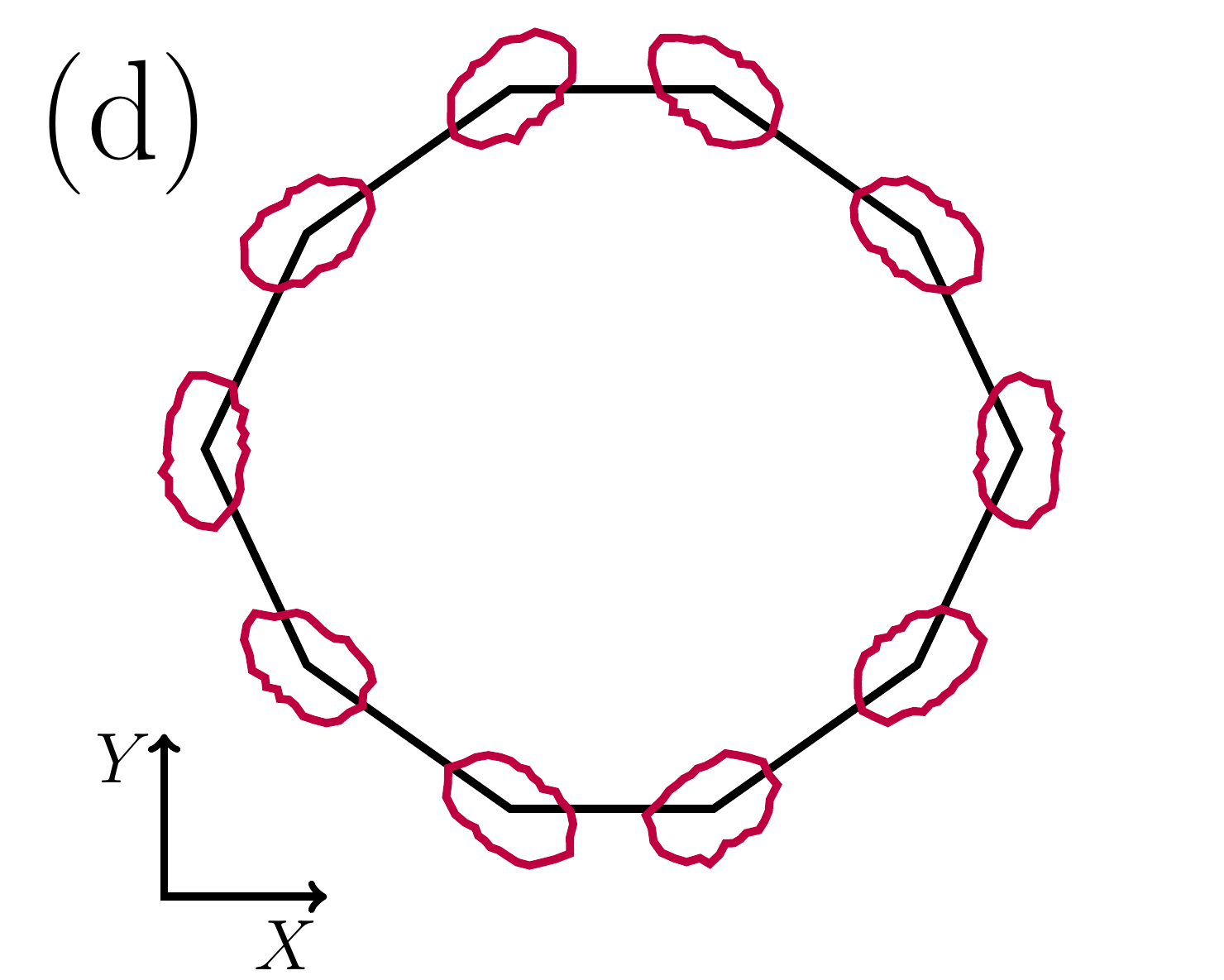}
	\caption{Probability distribution of the azimuthal angle for each spin in the coordinate frame of a 10-spin loop viewed from above. Panel (a) shows the distribution from earlier GGG refinements\cite{Paddison_2015_GGG}. Panels (b-d) show the distributions from the CNCS $E_{\rm i} =1.55$~meV, $E_{\rm i} =3.32$~meV and D7 refinements, respectively. Distance from spin to the surrounding contour is proportional to the probability for the spin of having the associated azimuthal angle.}\label{Richard:fig:phiDistribution}
\end{figure}

Fig.~\ref{Richard:fig:phiDistribution} shows the distribution of the azimuthal angle of the members in the loop in the coordinate system presented in Fig.~\ref{fig:YbGG}. We see that for YbGG, Fig.~\ref{Richard:fig:phiDistribution}(b-d), each spin is peaked along the tangent of the loop (local $z$-direction). This differs from the GGG refinements\cite{Paddison_2015_GGG}, where the distribution is peaked for angles perpendicular to the loop, Fig.~\ref{Richard:fig:phiDistribution}(a). 

The looped spin structure derived from the CNCS 3.32 meV dataset is more anisotropic than the D7 and the CNCS 1.55 meV datasets. It is estimated that this arises from a poorly sampled dataset, particularly for the medium to higher Q regions. Unlike the D7 and CNCS 1.55meV datasets, the CNCS 3.32 meV dataset does not have clearly defined features. The spin structure derived from this dataset is therefore less reliable.

\subsection{Notes on the D7 polarization}
 We have presented D7 data and corresponding simulations. The resultant spin-spin correlations and angular distribution show equivalence to those in the CNCS data, but the RMC fit is less convincing. There are several subtle differences between the CNCS and D7 neutron scattering intensities that may give rise to this. The CNCS magnetic scattering intensity, which we term $S_{magff}(Q)$ is obtained via the subtraction of a high temperature scattering from base temperature scattering. The high temperature scattering provides the intense magnetic formfactor and $S_{\rm magff}(Q)$ can result in negative intensities. This is considered in the RMC. D7 magnetic scattering, $S_{\text{mag}}(\textbf{Q})$,  is extracted using XYZ polarisation analysis with the following equation: 
\begin{equation}
S_{\text{mag}}(\textbf{Q}) = 2(I_{x,x'}+I_{y,y'}-2I_{z,z'})^{sf}, 
\end{equation}
in which $I_{x,x'}$ is the neutron $y$ spin flip scattering with the incident and scattered neutron polarisation along a Cartesian $x$ direction and $y$,$z$ denote the orthogonal directions,\cite{D7_Overview}. The resultant spin incoherent scattering is determined via: 
\begin{equation}
I_{\text{SI}} = \frac{3}{2}(-I_{x,x'}-I_{y,y'}+3I_{z,z'}). 
\end{equation}
The determination of $S_{\text{mag}}(\textbf{Q})$ in this manner assumes that the net moment of the compound is zero as is the case for paramagnetic systems or powdered antiferromagnet compounds and is thus employed for powder samples. A ferromagnetic signal would induce significant depolarisation of the scattered polarization. Using this equation for the case of a single crystal makes an implicit assumption that there is a net zero averaged moment with no symmetry breaking such that the magnetic cross-section is isotropic with magnetic components of equal magnitude projected along the three orthogonal directions. 
We made these assumption since we did not observe any depolarisation of the scattered beam, only short range order was observed and prior knowledge of the director state which provides an isotropic spin distribution, to a first approximation.  
Nevertheless $I_{\text{SI}}$ , expected to be homogeneous in \textbf{Q}, contains weak hexagonal features reminiscent of the magnetic signal. The peak positions of the spin incoherent signal are equivalent to the magnetic diffuse peaks in Fig.~\ref{fig:diffract}(b), and thus only peak intensities are affected while no shift of the peaks are observed. RMC optimises directly to $S(Q)$ and is sensitive to such relative changes. We suggest that these small variations give rise to the differences observed between the CNCS and D7 RMC and is the origin of the poorer simulations of the data. Nevertheless the resultant D7 RMC spin structure is consistent with that determined from the CNCS RMC and provides confidence in our results.
\section{A few notes on the Hamiltonian}\label{app:MC}
\subsection{Heisenberg model with anisotropy}

\begin{figure}[!h]
    \centering
	    \includegraphics[width=\linewidth]{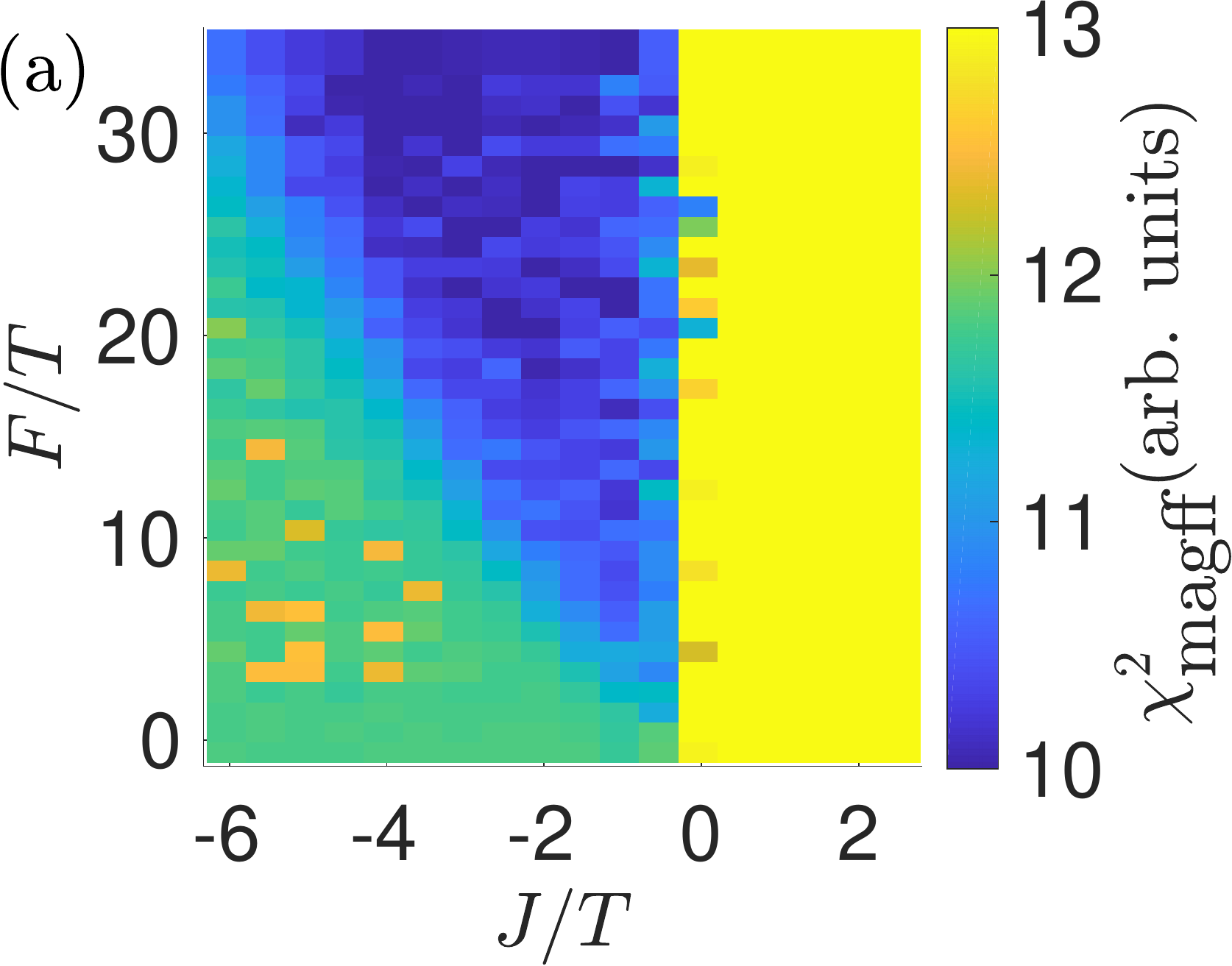}
	    \includegraphics[width=\linewidth]{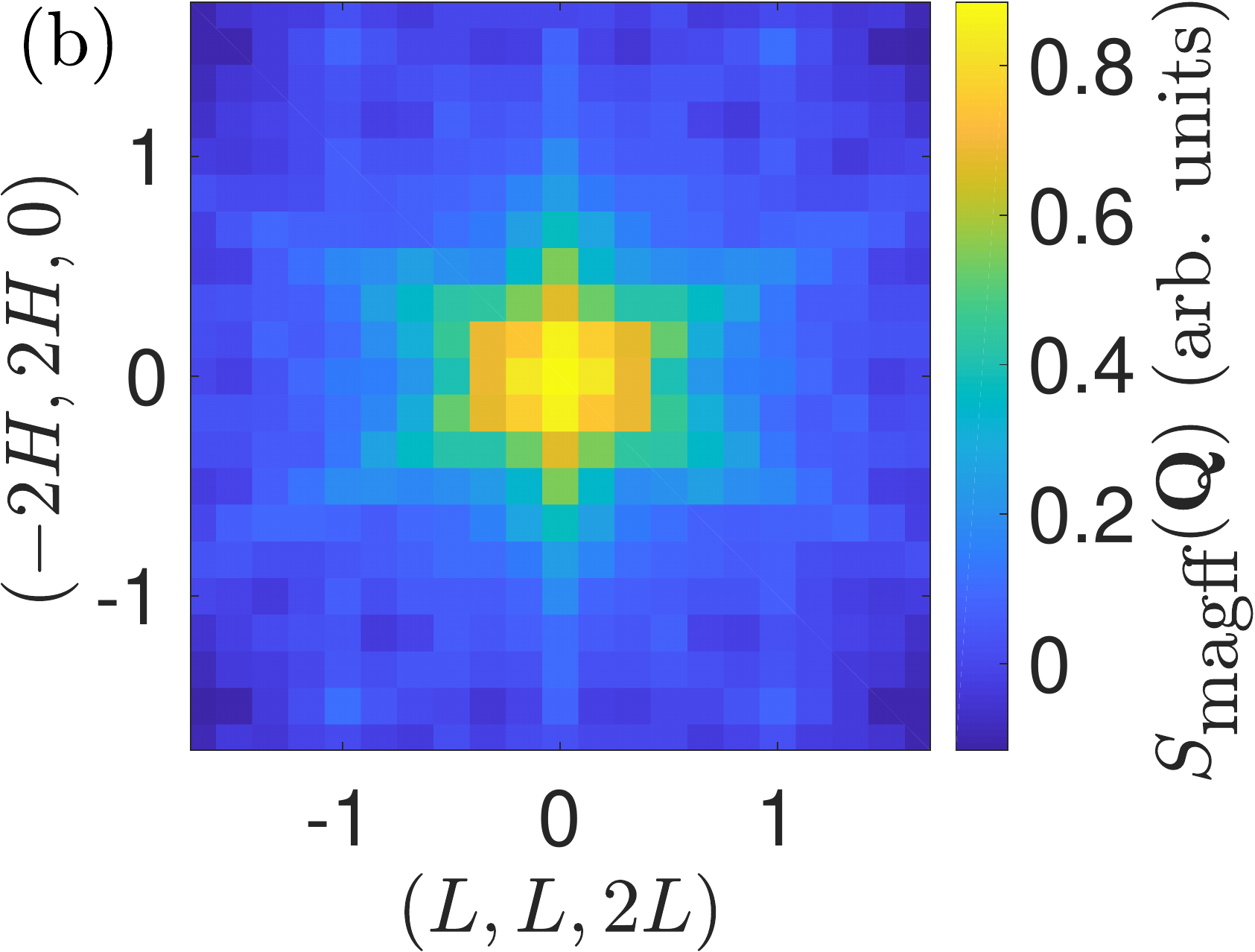}
	\caption{(a) Least square fit for the temperature reduced signal $\chi_{\rm magff}^2$ Eq.~(\ref{Richard:Eq:residualVector}). We vary $J/T$ and $F/T$ and calculate the residual for a system of $648$ ($L=3$) particles. The CNCS $E_{\rm i} =3.32$~meV data is used as reference. (b) Scattering profile for the best fit in this model. Here $J/T=-3,F/T=32$ (best fit in (a)) is shown for a system of $5184$ particles (L=6). }\label{Richard:fig:Heisenberg}
\end{figure}

The RMC method of the previous section suggests that spins have a preference to point along the tangential direction of the 10-spin loop. In particular, the distribution is peaked along the direction connecting the center points of two adjacent triangles, the local $z$-direction. Inspired by this result, we propose a nearest neighbour classical Heisenberg Hamiltonian with an energy penalty for spins pointing away from the axis direction,
\begin{equation}
    \mathcal{H}=J\sum_{\langle i,j\rangle} \textbf{S}_i\cdot\textbf{S}_j + F\sum_i |\textbf{S}_i-\textbf{S}_{i\parallel}|^2,
\end{equation}
where $\textbf{S}_{i \parallel}$ is the spin component along the local tangent axis (local $z$-direction) and $J$ is the strength of the nearest neighbour exchange interaction. In this simple Hamiltonian, $F>0$ models a classical easy-axis crystal field anisotropy and in the limit of large $F$, we obtain an Ising model. 
With the Metropolis-Hastings algortihm, we calculate a thermal average of the structure factor Eq.~(\ref{Richard:ScatteringLawEquationSystem}) and tune the parameters $J$ and $F$ to make the scattering pattern agree with the experimental data. Our best fit is shown in Fig.~\ref{Richard:fig:Heisenberg}(b). 
\begin{figure}[!h]
    \centering
    \includegraphics[width=0.3\textwidth]{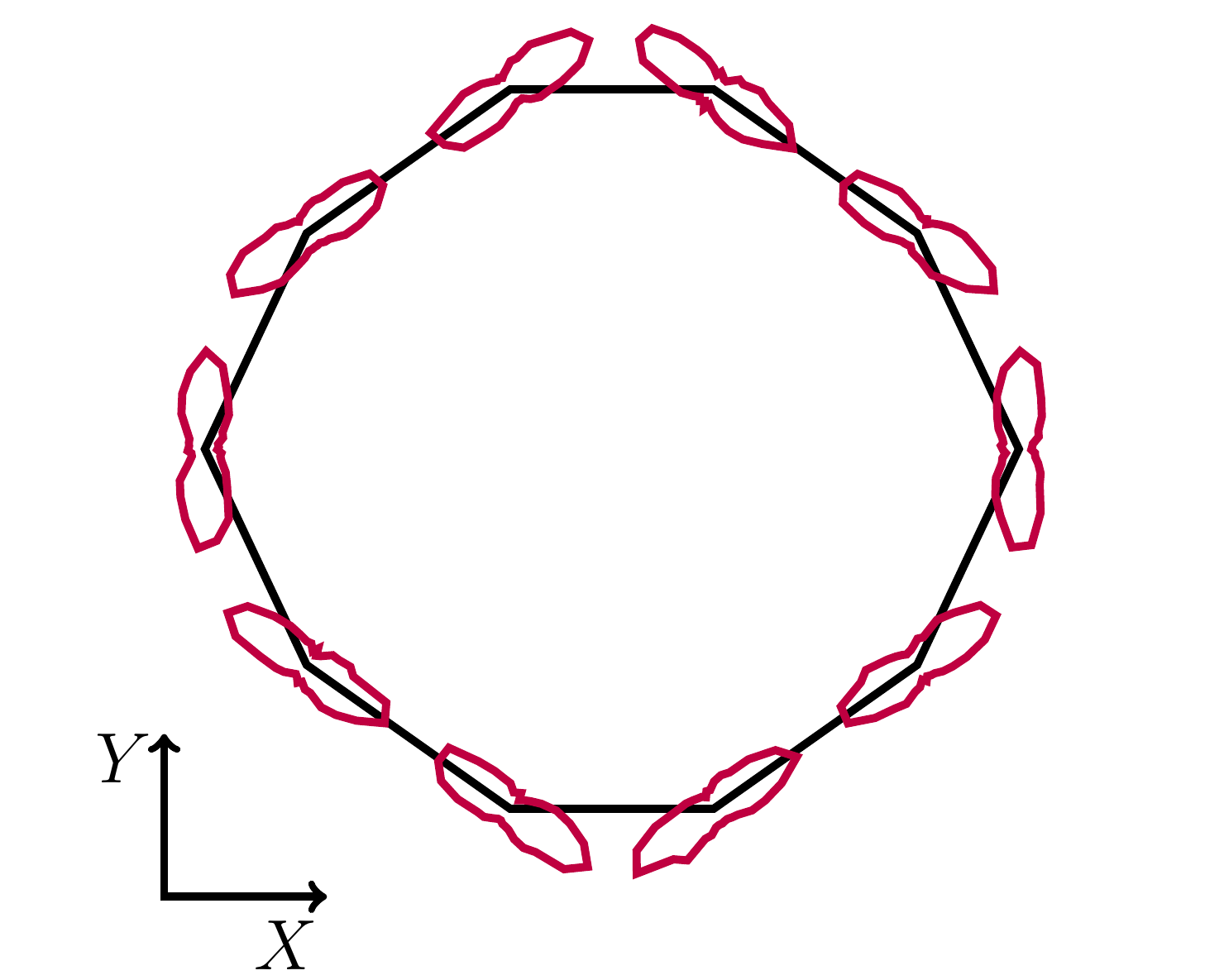}
    \caption{Characteristic Probability distribution of the azimuthal angle for each spin in the coordinate frame of a 10-spin loop viewed from above in the anisotropic Heisenberg model. Distance from spin to the surrounding contour is proportional to the probability for the spin of having the associated azimuthal angle.}
    \label{fig:Richard:HeisenbergPhiDistribution}
\end{figure}
In Fig.~\ref{Richard:fig:Heisenberg}(a) we also show the residual, $\chi^2$, Eq.~(\ref{Richard:Eq:residualVector}) with respect to the CNCS $E_{\rm i} =3.32$~meV scattering signal. Experimental data was binned to wavevectors periodic in the supercell. We show the error as a function of $J/T$ and $F/T$ for and 648 particles ($L=3$). We see that negative $J<0$ (FM NN interactions) give the best fit to the data. This is also in agreement with the Spinvert refinement that found a positive value for the nearest neighbour spin correlations, presented in the main text. From the parameter sweep, we see also that $\chi^2$ is minimized for large $F$. In the limit, we get an Ising model, which further motivates the crude Ising assumpton of the main text. We conclude by showing the characteristic spin distribution for the anisotropic Heisenberg model, Fig.~\ref{fig:Richard:HeisenbergPhiDistribution}.


	\centering

\section{Excitations}
\begin{figure}[!h]
	\centering
   \includegraphics[width = 0.9\linewidth]{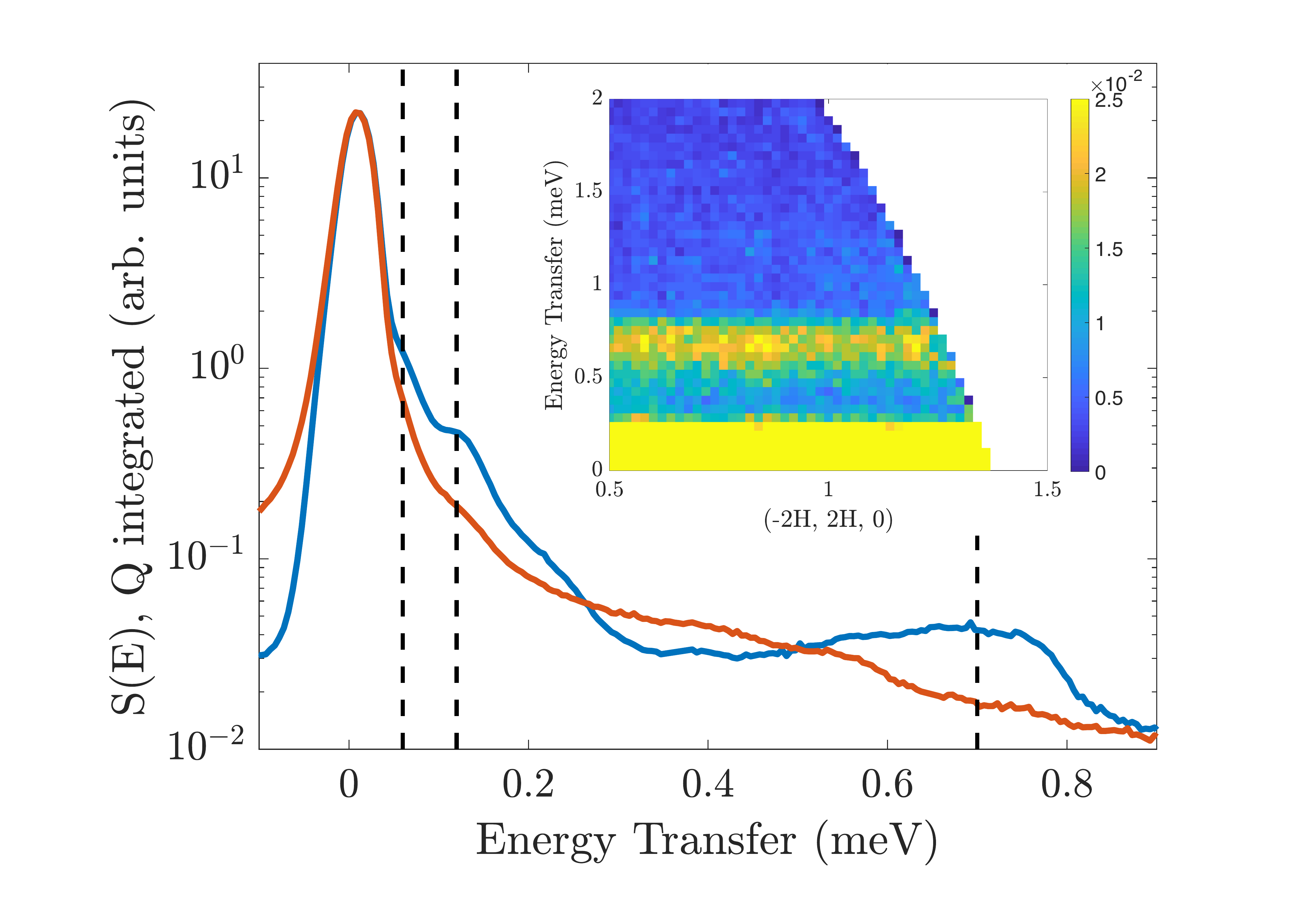}
	\caption{S(E), $E_{i}$ = 1.55 meV, 3 excitations, vertical dashed lines, are observed for the nominal temperature of 50 mK which are absent at 13 K.  (b): Cut in data with;  $E_{i}$ = 3.22 meV, along (L, L, 2L) with L = 0.23, clearly showing the dispersionless nature of the highest magnetic excitation.}
    \label{fig:sqw}
\end{figure}

Magnetic excitations have been identified within the CNCS data set. Three low lying dispersionless excitations are observed at 0.06, 0.12 and 0.7 meV at 0.05 K, see Fig. \ref{fig:sqw}that are absent at 13 K. The inset of Fig. \ref{fig:sqw} shows a  cut in CNCS data with incoming energy 3.32 meV with (L, L, 2L), L = 0.23, clearly showing the dispersionless nature of the highest magnetic excitation. A detailed analysis of these data will be published elsewhere.

\end{document}